# Where is my Glass Slipper?
# AI, Poetry and Art.

Dr Anastasios P. Pagiaslis

Assistant Professor in Marketing

Nottingham University Business School

University of Nottingham

Room B37, Business School North Building

Jubilee Campus

Wollaton Road

Nottingham, NG8 1BB

anastasios.pagiaslis@nottingham.ac.uk

# Abstract

The present literature review interrogates the multifaceted intersections between artificial intelligence, poetry, and art, offering a comprehensive exploration of both historical evolution and contemporary debates in digital creative practices. The review traces the development of computer-generated poetry from early template-based systems to advanced generative models, critically assessing evaluative frameworks such as adaptations of the Turing Test, the FACE model, and ProFTAP. The review examines how these frameworks endeavour to measure creativity, semantic coherence, and cultural relevance in AI-generated texts, whilst highlighting the persistent challenges in replicating the nuance of human poetic expression.

A distinctive contribution of this review is the integration of a Marketing Theory perspective. The Marketing Theory discussion deconstructs the figurative marketing narratives employed by AI companies, which utilise sanitised language and anthropomorphic metaphors to humanise their technologies. This critical examination reveals the reductive nature of such narratives and underscores the tension between algorithmic precision and the intricate realities of human creativity and experience.

The review also incorporates an auto-ethnographic account that offers a self-reflexive commentary on its own composition. By candidly acknowledging the use of AI in crafting this review, the auto-ethnographic account destabilises conventional notions of authorship and objectivity, resonating with Derridean deconstruction and challenging logocentric assumptions in academic discourse.

Ultimately, the review calls for a re-evaluation of creative processes that recognises the interdependence of technological innovation and human subjectivity. The review advocates for interdisciplinary dialogue that addresses ethical, cultural, sustainability and philosophical concerns in the age of AI, while reimagining the boundaries of artistic production and critique. This review contributes a nuanced perspective that bridges technical rigour with reflexive insight, inviting scholars to reconsider the evolving dynamics between human and machine creativity. By interrogating established paradigms and embracing its own hybridity, the review challenges readers to engage critically with the future of creative expression.

# Where is my Glass Slipper?
# AI, Poetry and Art.

## 1. Introduction

The following literature review discusses the latest papers on the role, challenges, ethical dilemmas, models and knowledge where Artificial Intelligence (AI) and Poetry intersect. To complete the review, I focused almost exclusively on articles that investigate the impact of AI on Poetry and not just art in general. I use articles regarding AI and Art sparingly to either show that results in Poetry have been corroborated in other areas or support certain topics like sustainability where evidence from poetry practice and AI does not yet exist. However, as Poetry is Art, I occasionally use the word 'Art' to denote that the observations of the review can be transferred to any artistic domain. The cut – off date for the articles is February 2024. Certain paragraphs of the present review lean on Nagl-Docekal and Zacharasiewicz's (2022) well curated volume on the relationship between AI and its potential for human enhancement and Linardaki's (2022) well referenced review.

This critical and comprehensive review contributes in the cross-disciplinary area of poetry and AI, or the domains termed as Digital Poetics (Weintraub and Correa-Díaz, 2023), Digital Humanities and possibly oddly enough Marketing Theory. This latter critique allows the present review to differ markedly from previous works because the review discusses the current literature from a Marketing Theory lens. The review may also contribute to the area of artificial intelligence, specifically focusing on the multifaceted aspects of poetry generation and its critical evaluation. The hope is that this review will serve as a valuable reference point for individuals who are either currently engaged in or are contemplating the pursuit of endeavours related to the generation of poetic works through AI technologies. Most of the AI systems the review discusses are still nascent and possess future research potential, particularly when one considers that there exist numerous AI technologies with alternative features, constraints, or advanced programming capabilities that current literature does not examine in depth. Most of these AI technologies may yield intriguing outcomes that not only contribute to the theoretical understanding of AI-generated poetry and Digital Poetics and all areas aforementioned but also offer potential practical applications in diverse fields such as education, entertainment and counselling. Ultimately, the intersection of AI and poetry generation represents a host of critical issues spanning philosophy to computer science and cognitive science to ethics, literature, marketing as well as a re-definition of creativity and innovation.

A final contribution of this review is the addition of an auto-ethnographic account in the end of the document on my own personal engagement with some of the systems the review reports on. The auto-ethnographical account operates as a vital 'supplement' to the ostensibly objective, critical discourse of the review. Derrida (1976) argues that meaning is never fixed; always deferred and shaped by traces of presence and absence. The auto-ethnographic account introduces a subjective, personal dimension that intentionally destabilises any claim to a pure, disembodied critique. Even in rigorous academic analysis, there is an inescapable imprint of personal experience — a trace that reminds us that all discourse is inherently interwoven with the lived and the particular, much like poetry. This insertion of a personal narrative challenges the logocentric assumption that meaning can be fully captured through detached critical analysis. Instead, the auto-ethnographic account reinforces the idea that texts (and by extension, experiences) are open to multiple interpretations, continually re-signified by the reader's engagement. In doing so, the auto-ethnographic account not only deepens the critical examination of AI-generated poetry but also reconfigures it, the review an intertextual conversation where the author's presence is both acknowledged and problematised.

The relationship between AI and poetry encompasses both technical and philosophical dimensions. Technologically, AI's involvement ranges from using machine learning algorithms for text classification to generating poetic content (Gatys, Ecker and Bethge, 2016). Natural Language Processing (NLP) techniques like Support Vector Machines, TF-IDF, and Doc2Vec analyse poetic texts and may enhance the understanding of stylistic nuances (Oliveira, 2017). Philosophically, AI-generated poetry challenges traditional notions of authorship and creativity, prompting a re-evaluation of what constitutes a poetic work, the role of the human poet and the definition of creativity (Chen, 2023).

The following sections attempt to engage with these issues discussing current technologies and narratives, technologies and frameworks for evaluating AI generated poetry, human perceptions of AI poetry, ethical concerns and bias, and Human-AI collaboration and offer a discussion of the current state from a marketing theory perspective. The final section provides my auto-ethnographic account. Throughout the review I use the phrase 'AI technologies' to speak about AI algorithms, models, and/or marketed systems in general and the phrase 'AI systems' when referring to specific examples of generative systems.

## 2. Poetry and AI History

Early NLP research in poetry generation, exemplified by the Bairon system, explored the use of templates and user input to create poems mimicking specific writers' styles (Karaban, & Karaban, 2024). AI poetry generation systems like ALAMO "rimbaudelaires" often rely on template-based approaches, using pre-defined structures with slots for specific words or phrases, to create poems (Uthus, Voitovich, Mical, & Kurzweil, 2019). However, Uthus et al. (2019) argue that these approaches can be restrictive and result in repetitive or formulaic poems, advocating for generative models that allow for greater diversity and quality of verses (Uthus and colleagues went on to develop Google's Verse by Verse system (Uthus et al., 2021)).

Examining the history of computer-generated poetry provides insights into the relationship between technology and creative expression, as well as the evolving capabilities and limitations of AI in replicating human-like creativity (Slater, 2023). Slater (2023) explores early Cold War attempts to generate poetry using computers, highlighting the influence of scientific, military, and corporate interests on these endeavours, noting that post-war computer poetry experiments often served to demonstrate the power and prestige of computers, with publicity stunts featuring computers generating poems used to promote the capabilities of these new technologies. Slater (2023) also explains that the development of computer-generated poetry also coincided with research in computational linguistics and machine translation, with poetry used as a test case to explore the linguistic processing powers of computers and these early experiments laid the groundwork for contemporary AI systems like GPT-4, which exhibit remarkable language generation abilities, including composing poetry that some describe as 'disturbingly humanlike'.

Scholars from different fields like engineering, psychology, mathematics, economics, and political science began exploring the creation of an artificial brain in the mid-20th century (Zulić, 2019). The official recognition of AI within the academic community occurred in 1956, with John McCarthy coining the term "artificial intelligence" in 1955 and that since then AI technologies are primarily designed to exhibit characteristics associated with human intelligence, such as understanding language, learning, reasoning, and problem-solving (Zulić, 2019).

d'Inverno & McCormack (2015) note that AI researchers are interested in building systems that can produce work considered "art" if created by a human, this goal embodies and refreshes the original ideas of AI as set out by McCarthy and colleagues in the 1950s. We can all of course recognise that over the past decade, the efforts of researchers in computational creativity have led to significant advancements in AI's ability to generate creative outputs, despite all the relevant and substantial problems associated with this rapid development.

Poetry was chosen as a field of exploration due to its concise form and rule-based nature, making it suitable for initial AI experiments with early attempts at poetry generation, dating back over seven decades, employing algorithms and pre-existing human-written poems (Linardaki, 2022). The first poetry generation program, "Stochastic Texts" (1959) by Theo Lutz, aimed to write poem-like texts from scratch in German and therefore even examples of early AI-generated poetry often exhibited grammaticality and some meaningfulness but struggled with poeticness, reflecting AI's limitations in understanding the nuances of language and creativity (Linardaki, 2022).

Even before Theo Lutz however Max Bense provides one of the earliest investigations of Artificial generation of Poetry (Beals, 2018). Bense used Shannon's information theory and suggested that aesthetic value could be measured quantitatively using statistical and formal methods. Then in collaboration with artists and engineers created some of the earliest interfaces for computer-generated poetry investigating the boundaries of subjectivity between humans and machines, representing a cyborg subjectivity (Beals, 2018). Bense's early experiments aimed to simulate human writing technologically and serve both as a thought experiment and an aesthetic practice. In Bense's works the human subject emerges through a stochastic process of automated text generation, representing a form of cyborg subjectivity. In this manner the final text becomes itself interface of human and machine, an intersection of multiple modes of textual production. As Beal (2018) notes Bense's work shows the merit of computer-generated texts in generating unconventional formulations. Bense himself distinguishes between 'artificial', 'synthetic', or 'technological poetry' and works by human authors. These early approaches already prompt reflection about the relationship between language and subjectivity, about the readers' urge to seek the self at the source of the text and about the roles of the author and subject in computer-generated poetry. For Bense the subject comes into being along with the text, hence the act of textual production constitutes a model of subjectivity. Bense's work influenced not only Theo Lutz but also the movement of the Algorists (see section 5.4 below). Beal (2018) notes that poetry was probably in the centre of conversations about AI, given its characterisation as a medium of subjectivity.

However, today the focus of AI poetry generation has shifted from solely generating grammatically correct and meaningful poems to exploring "intelligent" ways of generating poems that exhibit a higher level of creativity (Linardaki, 2022). Hence, researchers investigate how AI technologies can learn and apply complex computational and linguistic layers to produce novel and engaging poetic forms (Linardaki, 2022). Newer systems use of knowledge-intensive language generation models incorporates layers like phonetics, diction, syntax, and semantics relying on multiple algorithms and techniques such as: a) template generation or slot-filling, which involves using pre-defined templates with slots that are filled with words or phrases from different categories; b) Markov chains which use statistical models to predict the next word in a sequence based on the preceding words; c) recurrent neural networks a type of deep learning algorithm that can learn patterns in sequential data, making them suitable for tasks like language generation, d) Generative adversarial networks and e) Transformers (Linardaki, 2022).

As Linardaki (2022) notes a key challenge in AI poetry generation is enabling AI technologies to understand and generate meaningful and contextually relevant poetry that can capture cultural nuances and language polysemy. AI often struggles with meaning because it processes words without fully grasping their semantic depth leading to the need for the use of knowledge bases that provide AI with semantic information about words and their relationships (Linardaki, 2022). These knowledge bases can be created by analysing large corpora of text to identify patterns of word co-occurrence or by leveraging existing dictionaries and ontologies aiming to enhance the meaningfulness and coherence of AI-generated poems (Linardaki, 2022).

## 3. AI technologies and Public Narratives

The strides and public availability of generative AI is geometric, impacting various fields, including creative domains like poetry. AI can now generate poems in languages other than English, even languages that traditionally have been challenging for Large Language Models (LLMs) and Natural Language Processing (NLP) algorithms and most importantly limited public lexicon datasets exist, including for example Bangla (Murad & Rahman, 2023).

Karaban and Karaban (2024) explore the potential of AI-driven translation in poetry, comparing translations of Ivan Franko's poems by translator Percival Cundy and the GPT-3.5 language model suggesting that GPT-3.5 exhibits strengths in maintaining structural fidelity, rhythm, and rhyme schemes, comparable to human translation, although challenges remain in capturing subtle nuances of meaning and cultural context (Karaban and Karaban (2024). Alowedi & Al-Ahdal (2023) on the other hand note that current machine translation tools struggle to accurately capture the cultural context of art in this case that of Arabic poetry eschewing participants' appreciation of Arabic poetry and leading to translations that fail to convey the full meaning of the original text. Alowedi & Al-Ahdal (2023) note that human translation is generally superior for texts where 'accuracy, style, and cultural nuances are crucial'.

Li and Zhang (2020) show how a system using Natural Language Generator (NLG) algorithms can be combined with a Neural Machine Translation (NMT) technology to create poems aligned with a given theme, such as the phrase "Peach Blossom" in Chinese. Li and Zhang (2020) further note that from an algorithmic perspective the process of intelligent writing involves three major steps: ideation, material collection, and creative output. Li and Zhang's (2020) system draws on a range of techniques, including using text analysis technology to extract features like keywords, tags, sentiments and abstracts while also being able to trace the etymology of words, make connections between different data points and offer alternative forms of words in paraphrase. However, Phelan (2021:1) argues that such AI systems are not equivalent to any form of human literary critique and analysis, since the critical analysis of poetry is a separate activity from the creation of poetry and involves 'a sense of significance that is important'.

Hence, the challenge of imbuing AI technologies with an understanding of the nuances and complexities of human language remains at large. Amato, et al. (2019) agree and note that the complexity of replicating human creativity in poetry is due to factors such as imagery, cultural lexicons, and diverse poetic structures, necessitating sophisticated training and algorithmic advancements for AI.

Phelan (2021:1) exemplifies how *'the challenge for AI is to provide a database that recognizes what humans typically find humorous down to the subtle recognition of humorous tone'* while conceding that AI can draw on massive amounts of data to identify patterns in human language use, potentially leading to new insights and interpretations being able to detect secret sequences hidden from a less rigorous human analysis and make innovative connections of the kind made famous by AlphaGo's 'Move 37'. In every case, the artistic outputs of AI systems like 'Madeleine' (Phelan, 2021) challenge our understanding of the relationship between language and creativity, prompting us to reconsider the boundaries between human and machine intelligence.

Therefore, AI technologies and especially so generative systems force us to redefine the boundaries of poetry and creative outputs through the lens of technical innovation and philosophical inquiry (Oliveira, 2017). Notaro (2020) agrees, discussing the history of art engagement with computer art and generative art, focusing on AI-art and its financial and philosophical implications argues that AI art, or GAI-art, raises important philosophical questions, such as the meaning of being human in a technologically advanced world and the nature of creativity.

Then both AI research and implementations in the creative space require deep critical examination. Citizens and corporations should have the freedom to pursue their objectives, provided they do not infringe upon the rights or well-being of others (Nagl-Docekal and Zacharasiewicz, 2022). Regrettably though, public engagement in shaping research agendas remains limited, often confined to a consumer role (Knochel, & Sahara, 2022). Ethical considerations concerning risks, benefits, prohibitions, and permissions in AI are often delegated to expert commissions, parliamentary discussions, and government bodies and more often than not speculative narratives and the dominance of 'Big Tech' corporations shape public discourse on AI (Nagl-Docekal and Zacharasiewicz, 2022). These deliberations primarily focus on research funding and legal boundaries while 'Big Tech' corporations and their affluent 'celebrity' CEOs such as Elon Musk, Jeff Bezos, Bill Gates, and Jack Ma hold disproportionate sway over technological advancements, exceeding the influence envisioned by deliberative or discourse ethics approaches and with clear profit motives, consistently portraying their innovations as beneficial to attract consumers (Nagl-Docekal and Zacharasiewicz, 2022). More moderate voices like Jeoffrey Hinton raising concerns or even more so dissenting voices like Timnit Gebru become eventually silenced in the public and domineering narratives. This paradigm aligns with Marx's critique of capitalism, where economic growth, wealth accumulation, and private interests overshadow broader societal concerns (Nagl-Docekal and Zacharasiewicz, 2022).

On the other hand, public attempts to resist such narratives result in projects like the 'Vienna Manifesto on Digital Humanism' which calls for a comprehensive assessment of AI's capabilities and limitations, drawing insights from humanities, philosophy, theology, and theoretical physics emphasizing: the need to differentiate between strong and weak AI interpretations; the need to scrutinise practical applications; and the need to analyse AI's portrayal in fiction and film (Nagl-Docekal and Zacharasiewicz, 2022). While 'strong AI' attributes cognitive abilities and consciousness to machines, 'weak AI' views them as mere tools for specific tasks (Nagl-Docekal and Zacharasiewicz, 2022). Public apprehensions about AI often stem from speculative narratives usually grounded in pop-culture references when any discussions on agency and subjectivity involve a complex interplay of human and non-human actors (Nagl-Docekal and Zacharasiewicz, 2022). In every case, given the current levels of technology, fears of a "super-intelligence" arising from AI, though prevalent in science fiction, lack substantiation from current research (Nagl-Docekal and Zacharasiewicz, 2022). The singularity is not yet here despite the geometric advances in the AI space.

## 4. The Evaluation of AI generated Poetry

Pretsch (2023) argues that humans use poetry to communicate their feelings, observations, and experiences while AI analysing and pulling information from vast amounts of data creates poems by scanning thousands of poetic works across different styles and eras. This ability of AI prompts discussion about whether AI, without human experiences and emotions, can produce poetry that resonates with readers. Indeed Pretsch (2023) suggests AI poetry can evoke human emotions like sadness or happiness through themes of melancholy or joy. Despite AI's capability however the question if AI-generated poetry can be considered truly creative, given its reliance on algorithms and databases rather than personal experiences looms. Pretsch (2023:1) suggests that a counterargument to the criticism and questioning of whether AI can be truly considered creative is *'that AI's very detachment from personal experience might grant it a unique form of creativity. Without the biases, cultural impositions, and emotional boundaries that often constrict human poets, AI has the latitude to combine words, images, and sentiments in unforeseen ways'*.

The assertion by Pretsch (2023) brings back Nagl-Docekal and Zacharasiewicz's (2022) argument that the overemphasis on the speculative potential of "strong AI" can distract from the pressing ethical issues concerning the current and potential misuse of 'weak AI', particularly in data processing and bias replication. The rise of AI-powered creativity and expression support tools (CESTs) necessitates ethical guidelines that promote artistic expression while mitigating the risk of homogenization and the erasure of individual styles as well as the development and research on the evaluation of AI's actual Intelligence and inner workings and the 'creative' content such systems produce (Chung, 2022).

### 4.1. Formal Methods and Frameworks of Evaluating AI generated Poetry

A rigorous philosophical examination of non-instrumental reasoning is crucial to effectively guide the social, legal, consumer and personal regulation of digitalization processes (Nagl-Docekal and Zacharasiewicz, 2022). Human understanding, characterised by its intersubjective dimension, cannot be fully replicated by algorithms, even those capable of "learning" and engaging in rudimentary communication (Nagl-Docekal and Zacharasiewicz, 2022). The claim that a reductionist, instrumental approach can adequately explain human comprehension and interaction rests on an unfounded belief in scientism (Nagl-Docekal and Zacharasiewicz, 2022). Human understanding encompasses various modes of non-calculative reasoning, including artistic expression and religious interpretation, that defy capture by purely instrumental rationality (Nagl-Docekal and Zacharasiewicz, 2022).

Prominent tests designed to evaluate a machine's ability to mimic human-like intelligent behaviour like the Turing Test that for decades has been a prominent benchmark in AI research are no longer enough (Nagl-Docekal and Zacharasiewicz, 2022). Clements (2016) argues that relying solely on mimicking human language, as exemplified by the Turing Test, may lead to unoriginal and clichéd poetry. For that matter the literature presents alternative models of AI generated poetry although still in early phases of development. Clements (2016) discusses various algorithms of Poetry production from Markov Chain algorithms to smartphone apps such as Swift-speare an app which generates sonnets and was used in 2014 to produce a sonnet and generate some discussion in media via publication in TechCrunch.

Clements (2016) argues that evaluation of machine learning algorithms is now looking into an adaptation of the Turing Test as a 'Poetry Turing Test' which checks for a cultural match based on accepted poetics and then attempts to emulate traditional verse forms and the poetic styles of established poets potentially evoking the eloquent, speaking subject. The caveat is of course that to the day systems like the Swift-speare app still require significant human intervention and input. Matias (2014 in Clements, 2016) states that the sonnet that was compared to Shakespeare's original sonnets was essentially written in collaboration with *suggestions* from the app. Clements (2016) provides another interesting example of Professor Parker to evaluate the preference and discernment of AI generated poetry from a lay audience by generating a blind in-between subjects experiment between a Shakespearean sonnet and a computer-generated sonnet; surprisingly or perhaps not, most participants preferred the computer-generated sonnet, although Parker himself acknowledged that the Shakespearean sonnet was likely the superior work of art.

Deng, Yang, & Wang, (2024) propose ProFTAP as an alternative algorithmic framework that analyses how distinguishable AI-generated poetry is from human-generated poetry, and involves four steps: data collection and pre-processing, AI preparation, generation and post-processing/anti-plagiarism, allegedly these four steps aim to make the framework more objective and rigorous in assessing AI's poetry writing capability compared to previous methods. According to Deng, Yang, & Wang, (2024) finetuned open-source LLMs can generate classical Chinese poems nearly indistinguishable from those of ancient Chinese poets. However, worth noting that the ProFTAP framework employs the use of Human judges as arbitrators of distinguishability essentially eventually calculating an inter-judge reliability metric.

Oliveira (2017) expands on yet another model that aims to evaluate AI generated poetry: the FACE model. Colton, Charnley, & Pease (2011; see also Colton & Wiggins, 2012; Colton, Goodwin, & Veale, 2012) developed the FACE framework as a descriptive model that AIs can use to evaluate if the poetry AIs create is comparable to human poetry. FACE assesses a poem or for that matter a creative system positively by asking and expecting the system to *'create a concept (C), with several examples (E), include an aesthetic measure (A) for evaluating the concept and its examples, and provide framing information (F) that will explain the context or motivation of the outputs'* (Oliveira, 2017: 18). However, as Oliveira (2017) points out the act of framing, which can indeed be presented in the form of a commentary articulated through natural language, does not significantly alter or impact the evaluations made by humans regarding the creativity, significance, or overall quality of poems that are generated by computers.

Wu, Song, Sakai, Cheng, Xie, & Lin (2019) discussing the evaluation of poetry inspired by specific images and produced by Recurrent Neural Network (RNN) generators note that two challenges stand out: First, evaluating generated text without ground truths. Second, evaluating nondeterministic systems that may produce different texts from the same input image. Wu et al. (2019) suggest that in such case traditional metrics like the Bilingual Evaluation Understudy (BLEU) are less suitable due to the lack of ground truth and the importance of creativity hence composite measures of quality, novelty and diversity are needed. In Wu et al.'s (2019) work human assessors rate quality while novelty proxying Boden's H-Creativity and diversity proxying P-creativity are algorithmically computed by assessing the presence of new k-grams in the generated poem compared to a training corpus.

Finally, Tanasescu, Kesarwani, & Inkpen (2018) suggest that techniques like Convolutional Neural Networks (CNN) and deep learning approaches provide promising results in metaphor detection and the automatic analysis of poetry. All in all, the formal evaluation of AI generated poetry is as difficult as the evaluation of human created poetry. The challenge remains in the pervasive inability of automated and non-automated frameworks to assess aspects like grammaticality, meaningfulness, and poeticness of a generated poem.

### 4.2. Human Perceptions of AI generated Poetry

Formal models and frameworks notwithstanding seems to be the case that the general public as aforementioned in the experiment run by Professor Parker (Clements, 2016) not only cannot distinguish necessarily between AI generated and Human created poems but also has a rather unstable preference for AI/Human generated poems and art. Köbis & Mossink (2021) echo Professor Parker's experiment and with a similar in-between subjects experiment show that people cannot differentiate AI-generated from human-written poetry although their results contrast with Professor Parker's regarding preference. Köbis & Mossink's (2021) studies are extremely interesting because they show that although respondents clearly state an aversion or dislike to algorithms executing emotional (vs. mechanical) tasks as a general attitude these views correlate consistently, but weakly (b = -0.16, p = 0.13; although in a different statistical tradition the p-value for the regression coefficient would be considered marginally insignificant), with respondent behaviour in choosing human poems over algorithmic poems. Köbis & Mossink (2021) also show the importance human intervention still has in regulating AI generated poetry. Only poems selected by the experimenters successfully passed as human and lowered algorithm aversion while in all other cases this aversive tendency did not change regardless of respondents being informed about the algorithmic origin of the text even when disclosed before exposure to the AI generated poem. Which points to the fact that despite the generalised aversive attitude, when people are presented with the final output their evaluation of said output is almost independent from their general attitude.

Hitsuwari, Ueda, Yun, & Nomura (2023) further investigate the emotional connection people form with art and reveal that knowledge of an artwork's origin, whether human or AI-generated, influences emotional responses. Hitsuwari, Ueda, Yun, & Nomura's (2023) results corroborate the aforementioned studies and show that haiku poems generated through human-AI collaboration received the highest beauty ratings. In contrast, the beauty ratings for human-made haiku and AI-generated haiku without human intervention were equal. Participants in the study were, as previously, unable to reliably distinguish between haiku poems created by humans and those generated by AI.

Badura, Lampert, & Dreżewski (2022) show that using a specialty poetry trained model based on existing AI technologies and established poets' poems may produce without human intervention indistinguishable verses compared to the original poets. Badura, Lampert, & Dreżewski (2022) however evaluate not a complete AI generated poem but single verses on general quality, grammatical correctness, and resemblance to the poet's literary style using these three criteria as a proxy to poem quality. Then combining subjective human ratings and a formal automated evaluation framework including the Bilingual Evaluation Understudy (BLEU) score, emotion analysis and word percentages show that their trained model was able to produce texts that are difficult to distinguish from the original works on a line-by-line basis.

Demmer, Kühnapfel, Fingerhut, & Pelowski (2023) investigate the impact of artist provenance on viewers' emotional responses to art, as well as their tendency to attribute intentionality to the artist. Their study specifically examines how knowledge of whether a work of art was created by a human or an AI influences these two types of responses (Demmer, et al., 2023). Using a 2 x 2 within-participant design Demmer, et al. 2023) present simple forms of artwork with preceding information about whether the artwork was made by a Human or an AI with said information being accurate in 50% of the cases, examining the impact of both actual provenance as well as the effect of information communication. The study's results interestingly indicate that participants did report feeling emotions when viewing AI-generated art and frequently reported that they believe the computer intended to evoke specific emotions; at the same time participants report stronger emotions when viewing art they believe was created by a Human, even when this assertion was merely false information (Demmer, et al., 2023). While this suggests AI art can evoke emotional responses, human-made art elicited stronger emotions, highlighting the potential influence of perceived artist provenance on audience engagement (Demmer et al., 2023).

In a similar vein, Mikalonytė and Kneer (2022) explore whether people are willing to consider paintings made by AI-driven robots as art and robots as artists. Mikalonytė and Kneer (2022) run 2 separate pre-registered experiments testing respondent perceptions on abstract vs. representational painting. Both experiments feature 2 (agent type (AI-driven robot vs. human agent)) x 2(behaviour type (intentional creation vs. accidental creation)) between subjects design. Mikalonytė and Kneer's (2022) results show that respondents rate both robot and human paintings as art to roughly the same extent. However, respondents correlating perceived mental states with artistic agency are less willing to consider robots as artists than humans, since the former do not possess artistic intentions. Hence, the interaction between the agent (e.g., an autonomous robot, as compared to human), the process (the action by which a work is brought to life), and the product (the created object) influences the formation of respondent perceptions.

Gunser, Gottschling, Brucker, Richter, Çakir, & Gerjets (2022) examine how reader perceptions and differentiation between AI-generated versus Human-written poems influences quality perceptions. Participants were asked to classify text continuations as AI-generated or Human-written and to rate their stylistic quality (Gunser et al., 2022). Their results show that participants were not only generally inaccurate in differentiating between the AI-generated and Human-written continuations but also overconfident in their decisions while perceiving AI continuations as significantly less well-written, inspiring, fascinating, interesting, and aesthetic than Human-written continuations. Regarding prior knowledge of provenance Gunser et al. (2022: 1746) note that *'the proportion of AI-generated and Human-written texts was undisclosed to ensure that decisions could not be derived based on previous trials'*.

Horton, White, & Iyengar (2023) show that in the visual arts algorithmic aversion against AI art can enhance perceptions of human creativity, suggesting that social and cultural factors play a significant role in shaping our appreciation of art. Horton, White, & Iyengar's (2023) results through multiple studies and with a combination of within and in-between subjects experiments show that participants exhibited as in previous studies a general attitudinal aversion against art labelled as AI-made. Participants evaluated these images less favourably across various dimensions, even when participants, as previously, were incapable of distinguishing AI-made art from Human-made art. Particularly participants rated expensiveness and skill of AI generated images less favourably. Participants however confirmed that AI-labelled images qualify as art. Finally, an extremely interesting result of Horton, White, & Iyengar (2023) is that their studies also showed that direct comparisons can enhance perceptions of human creativity, with human-labelled art being rated more favourably when compared to AI-labelled art especially so when Human-labelled images followed AI-labelled images.

Formal models/frameworks and general public's evaluation of AI generated poetry notwithstanding one needs to accept that the utilization of extensive statistical knowledge bases and distributional models of semantics has enabled AI technologies to produce text exhibiting semantic coherence, albeit raising questions about their genuine understanding of meaning (Nagl-Docekal and Zacharasiewicz, 2022). The ability of AI technologies to generate creative outputs, like poetry and art, raises questions about the nature of creativity and the role of AI in artistic practices (Köbis and Mossink, 2021). Abramson (2021) suggests that from a cognitive science perspective Explainable Computational Creativity (XCC), a branch of Explainable AI (XAI), is crucial for understanding how AI produces creative works and addressing concerns about its autonomy and potential misuse. Regardless of the technology or criteria society decides to use to evaluate AI-generated poetry, AI's potential to create aesthetically pleasing and semantically plausible work relies on simultaneous technological progress in artificial general intelligence (AGI) and as well as usually strict adherence to genre traditional and occasionally stereotypical conventions. Manovich (2018) discussing AI aesthetics asserts that while simple AI can generate satisfying results in certain genres, particularly those with looser conventions like experimental poetry, replicating the complexity traditional poetry requires a tighter coordination between elements and deeper understanding of cultural nuances. Manovich (2018) further notes that current systems mainly work (especially so quantitative algorithms) on the basis of using first shared traits and commonalities and second differences and uniqueness and / or originality to categorise cultural artefacts. As AI acts like a cultural theorist by identifying common patterns in culture and evolves and influences cultural lives and aesthetic choices (remember AI decides what pictures and newsfeed you will receive), automation raises concerns about potential decreases in aesthetic diversity. Therefore Manovich (2018) proposes that a potential framework against reduction of diversity may lie in the development of a 'Cultural Analytics' field of knowledge that aims to visualise all artifacts without oversimplification, focusing on differences between cultural artefacts rather than shared traits while recognising that uniqueness in cultural artefacts is essential for understanding diversity.

The ongoing debate surrounding AI-generated poetry exemplifies the difficulty in defining both poetry and art as objects themselves, artificial intelligence, as well as areas traditionally considered the pervasive fields of human functioning such as creativity and the creative endeavour. Such a discussion prompts considerations about the role of ambiguity, formal and semantic openness, and the interpretive community in evaluating poetic works (Köbis and Mossink, 2021). The ability to evoke ambiguity is often regarded as a uniquely human attribute, and while AI technologies can be trained to recognise irony and ambiguity to a certain extent, they struggle to achieve the subtlety and nuance characteristic of human poetry at least yet (Miller, 2019). Current AI poetry generation systems employing various techniques, such as linguistic templates, evolutionary algorithms, and neural networks, to create poetic content are still guided by specific constraints like metre, rhyme, and semantic relatedness to produce more structured and coherent poems (Misztal & Indurkhya, 2014).

Grba (2020) argues that 'the poetic realm of contemporary AI art is most deficient in interesting intuitions, meaningful abstractions, and imaginative analogies. The field particularly lacks projects that use AI as means to actualise strong concepts which effectively address the wider perspectives or deeper issues of human existence' (Grba, 2020: 49). On the other hand, Epstein et al., (2023:1) emphatically assert that the development of AI systems like ChatGPT suggests that AI can produce *'high-quality artistic media for visual arts, concept art, music, fiction, literature, and video/animation'*.

## 5. Algorithmic Bias and Ethics in AI Generated Poetry

West & Burbano (2020) editors of the special issue of Artnodes on AI, Arts and Design note that ethical considerations in AI Poetry and art production need also be a significant point of discussion. Galanter (2020) contributing to that discussion introduces the concept of 'machine patiency', advocating for ethical treatment of machines that create art, especially as they exhibit increasing autonomy in their creative processes. Galanter (2020) suggests that as AI becomes more sophisticated in its ability to learn and create independently, moral considerations become crucial.

Vyas (2024) suggests that AI poetry and art ethics should consider disinformation, the production of poor-quality content, and barriers to communication between stakeholders. A common example of troubling content for instance is AI-generated text which while well-composed, might misattribute theories or concepts, potentially misleading readers.

### 5.1. Will the artists survive?

d'Inverno & McCormack (2015) suggest that the development of AI technologies that can generate creative outputs, including poetry, raises ethical questions about the nature of creativity and the potential impact of AI on human artists, since some researchers and artists focus on AI as a tool that can augment and enhance human creativity (focus on the 'weak AI' narrative – see section 3), while others express concerns about the potential for AI to replace human artists (currently unsubstantiated fears – see section 3). Fathoni (2023) correctly points out that the increasing availability of AI-based art models, such as Dall-E and Midjourney, has sparked debates about the role of AI in art and the implications for the art market and the livelihood of artists These livelihood concerns echo the issues faced by artisans during the Industrial Revolution, where the introduction of new technologies led to dramatic and painful changes in the nature of work and the value of human craftsmanship occasionally leading to significant conflicts (Newton & Dhole, 2023).

Epstein et al. (2023b) resolutely assert that the disruptive potential of AI may indeed displace some existing jobs in the creative industry but will surely also create new opportunities for creative labour. Jordan (2020) echoes this assertion and suggests that jobs involving data compilation and interpretation are at higher risk of replacement by AI than artists, with jobs in the arts and entertainment being amongst the least vulnerable to automation, as they require personal skills and creative abilities that AI currently struggles to replicate. Interestingly enough, Jordan (2020) is a member of the Arts Management & Technology Laboratory at Carnegie Mellon University.

As such creative AI seems to have an immense impact on creative communities, but along with great power comes significant ethical responsibilities in AI's set up, use, and in the output works themselves say Flick & Worrall (2022) echoing our beloved neighbourhood Spiderman. Generative AI (GAI)-poetry and art raises major philosophical questions including the meaning to be human in a hyper-connected world and the true nature of human creativity (Notaro, 2020). While its application generation in creative domains raises questions regarding the credibility of AI-generated content (Gunser et al., 2022). Hence, the potential of AI poetry and art as an autonomous genre raises issues for artists and experts surrounding the use of AI systems in creative practice with the central question arising: Is or Should AI-generated 'Poetry/Art' by means of using algorithms and extensive public or otherwise datasets be considered as true art? The question remains unanswered.

## 5.2. Algorithmic and Dataset Bias

The use of large datasets in LLM and AI training raises multiple concerns about algorithmic bias, as AI technologies tend to reflect and amplify existing biases present in the data they are trained on and of course existing biases of the engineers that wrote, designed and compiled the system (Elam, 2023). AI's tendency towards 'algorithmic ahistoricity', stemming from its reliance on uncritically fed available 'public' (i.e. Internet) historical data, can lead to the generation of poems that reinforce harmful stereotypes or perpetuate social injustices (Elam, 2023). Additional to the concerns about perpetuating and possibly reinforcing already existing human biases the ethical concern about the commodification of what was until recently considered a public domain, that of the Internet as well as the works that for better or worse belong now to this 'public' domain remains looming and brings forward questions of authorship, intellectual property rights, the essence of the poetic experience and 'the commodification of the human spirit by mechanising the imagination' (Cave, 2023; Chen, 2023).

Epstein et al. (2023) emphatically assert that technological advancements drive the progress of AI in creative fields with diffusion models enabling the synthesis of high-quality images, and large language models (LLMs) facilitate the generation of impressive prose and verse, and these advancements empower both artists and AI technologies to explore new creative possibilities. However, although the availability of large-scale training datasets and increased computing power, particularly through GPUs, plays a pivotal role in enabling AI technologies to learn and improve their creative capabilities this unfettered access to data, pours oil in the fire of concerns about potential bias in AI-generated content (West & Burbano, 2020).

Miller (2019) suggests that the advent of AI generated poetry, and art introduces a new realm of technical practices and interpretive frameworks for creating machines exhibiting intelligent behaviour. Artists have already – both silently and openly – embraced AI practices, integrating them into their creative processes, resulting in what is often termed 'Expressive AI' (Miller, 2019: 64). Here, AI technologies act as artistic artifacts (or tools – weak AI), enabling artists to convey complex ideas and experiences to their audiences (Miller, 2019). Expressive AI perceives AI as a performative medium through which the artist's intentions are communicated (Miller, 2019).

However, despite the adoption of AI technologies by artists either silent or open the production of AI generated or AI/Human collaboration generated poetry, and art abounds biases and ethical quandaries. Srinivasan and Uchino (2021) note that examining AI art through the lens of art history reveals biases inherent in the data used to train AI models, potentially replicating historical inequalities and limiting the diversity of artistic output. A critical understanding of these biases is essential to ensure that AI art promotes inclusivity and avoids perpetuating harmful stereotypes (Srinivasan and Uchino, 2021). On the other hand, Cetinic & She (2022) suggest that AI technologies enable new ways to analyse and visualise digitised art collections due to advanced computational methods that can enhance research perspectives. However, Cetinic & She (2022) focus solely on the 'weak AI' narrative discussing how automatic classification of artworks through Convolutional Neural Networks (CNNs) increases object identification and similarity retrieval while allowing for multi-modal tasks hence potentially allowing for enhanced knowledge discovery in art history suggesting that 'Digital Art History' is emerging as a significant field of study. Cetinic & She (2022) do though recognise that many AI artworks rely on human curation and input for their creative process while the narrative of autonomous AI artists is often driven by marketing claims while also recognising that questions of authorship and profit arise with AI-generated artworks and that copyright issues are complex due to the use of existing copyrighted images in training.

Tromble (2020) argues that AI entities, including robots, Gen-AI and AI algorithms in processes like feeds of various social media platforms, perpetuate existing power structures, particularly in tasks that differentiate human from non-human. Mateas (2001) twenty years before the advent of market availability of AI highlights the potential of AI to engage audiences in the exploration of complex meanings through art although notes that AI researchers often prioritise task competence and demonstrably accomplishing well-defined tasks as such this focus on objective measurement and general principles contrasts with artistic goals of conveying nuanced meanings and exploring specific cultural contexts. Flick & Worral (2023) set out the key ethical issues relating to creative AI as copyright, replacement of authors/artists, bias in datasets, artistic essence, dangerous creations, deepfakes, and physical safety.

**5.3.    AI Anthropomorphising**

As AI research increasingly focuses on developing technologies that exhibit intelligent behaviour, blurring the lines between tool and creator, the anthropomorphic depiction of AI, often in the form of feminized avatars raises ethical questions especially when such depictions can reinforce gender stereotypes (Chused, 2023). The perception of AI technologies as more trustworthy or empathetic based solely on appearance should be critically examined considering feminist critiques and discussions on the societal impact of AI (Nagl-Docekal and Zacharasiewicz, 2022). The use of anthropomorphic language and imagery in AI presentations can lead to a distorted view of AI technologies as autonomous entities possessing human-like intentions and emotions (Nagl-Docekal and Zacharasiewicz, 2022).

Although many professional artists have voiced general concerns about the harms they have experienced due to large-scale image and text generators trained on image/text pairs from the internet, including reputational damage, economic loss, plagiarism, and copyright infringement, there are concerns that anthropomorphising image and text generators further diminishes human creativity, robs artists of credit and compensation, and ascribes accountability to AI systems rather than those who create the systems and those who create the data systems feed on (Jiang et al., 2023).

Equally ethically questionable is the attempt to imbue poetry generation technologies with emotion (e.g. Kirke & Miranda, 2013). Kirke & Miranda (2013) introduce MASTER, a poetry generator that uses a multi-agent system and digital emotion to analyse texts for emotional content and then generate poems that reflect the perceived emotional state of the system. These so called 'agents' are not live people, but rather cyber entities / 'autonomous digital entities' designed to interact within the MASTER system. These agents have digital emotional states and attempt to influence each other's emotions by reciting "poems" to one another, reflecting their own emotional states. The agents operate within a simulated environment, adjusting their emotional states and recited texts based on interactions with other agents, rather than being influenced by human emotions or actions. Kirke & Miranda (2013) suggest that the design of these agents is intended to explore the dynamics of emotional influence and creativity in a digital setting, rather than involving human participants directly. As such Kirke & Miranda (2013) suggest that the MASTER system can generate poems with emotional depth and thus opens up new possibilities for creating poems that express specific moods or evoke particular emotions in the human reader.

If AI technologies eventually are represented by anthropomorphic avatars cultivating their own emotions via some agentic system what is the role and nature of the poetry produced and what is the role of the poet? Where will human poets find their place in such a reality? Hutson & Schnellmann (2023) argue that concerns regarding the "death of the artist" due to AI are unfounded, highlighting historical precedent for the adaptation of new technologies by artists and poets and suggesting that AI's role is more likely to be collaborative than directly replace the poets and artists. Spencer (2023) on the other hand expresses a contrasting perspective, arguing that AI poses a threat to the human imagination, potentially limiting human self-actualisation and artistic expression.

Notaro (2020) notes that AI artist, Mario Klingemann, suggests that AI is about augmenting the human imagination. However, tools such as OpenAI's GPT-3, that have written newspaper articles, conference talk titles, and creative fiction, highlight the concern that creatives could be replaced through clever use of creative AI (Notaro, 2020). A sentiment echoed partially by other AI artists as discussed in section 6.

## 5.4. Creativity, Authorship, Ownership, Copyrights and Intellectual Property Rights

Brown et al. (2020) reinforce and assert the notion that the growing sophistication of large language models (LLMs) like GPT-4 and BERT raises new challenges and opportunities for creative practices, prompting a reassessment of traditional notions of authorship, originality, and artistic agency. The ease with which LLMs can generate text that mimics human writing styles necessitates a critical examination of the implications for artistic integrity and the potential for misuse (Kobierski, 2023)

However, and despite the rapid advancements in AI poetry generation, researchers acknowledge that current systems are still a work in progress. Kirmani (2022) in his viewpoint paper acknowledging this state suggests that the challenge for current technologies lies in moving beyond mimicking existing patterns to achieving genuine creativity and originality, although provides no definition for the latter concepts. On the other hand, Miller (2019) adds that understanding how AI technologies create poetry may provide insights into the nature of creativity, challenging our assumptions about human exceptionalism and prompting us to re-evaluate the boundaries between human and machine. Therefore, one needs to consider evolving definitions of creativity from different lenses and narratives ranging from creativity as a re-combinatory and evolutionary phenomenon to emphasizing novelty, originality and technology.

Following through Flick and Worral (2023) argue that an ethical creative AI would incorporate the user into the workflow, rather than just incorporating the user's identity/essence/data. Similarly, Hassine & Neeman (2019) contend that AI is not a major artistic breakthrough in its current form as AI poetry and art projects reproduce artists styles without their creative input in fact AI-generated works should be considered a form of artistic plagiarism because AI-generated poetry and art simply replicates stylistic elements without the artist's permission. Hassine & Neeman (2019) propose that aesthetic evaluation should be based on creativity, innovation, and a sense of surprise. In contrast, Barale (2021) argues that AI art is inextricably linked to aesthetics, defined as our sensory experience of the world. Because the machine must 'learn to see' and share its vision with the human artist, AI art offers a unique perspective on our perception and representation of the world.

Adding to the debate, Jiang et al. (2023) argue that image and text generators are not artists. For Jiang et al. (2023) art is a uniquely human endeavour connected to human culture and experience, drawing on philosophies of art and aesthetics. Art uses cultural resources to embody experience in a form that is accessible to an audience. This process requires control and sensitivity to the perceiver's attitude, distinguishing it from spontaneous activities driven by organic pressures. The human element transforms these activities into those performed to elicit a response from the audience. As such and to the day image and text AI systems require human aims and purposes to direct their production, and these human intentions shape their outputs (Jiang et al., 2023).

Together with creativity the fields of poetry, literature and the arts will be forced to re-define ownership, copyright and intellectual property rights of AI-generated poetry. Who owns the rights to a poem created by an AI? Is it the creator of the AI technology, the user who provided the input/prompt or the AI itself? Or maybe the artists whose data were used to train the system? These questions remain largely unanswered, and legal frameworks are not keeping pace with the rapid development of AI technology.

A related concern is the use specifically of copyrighted works in training AI models. Some question whether this constitutes infringement, and the lack of clarity in copyright laws regarding fair use in the context of AI adds to the complexity of the issue (Piskopani, Chamberlain, & Ten Holter, 2023). Degli Esposti, Lagioia, & Sartor (2020) explore automatic methods that quantify the similarity between AI-generated works and existing copyrighted material, drawing on concepts from information theory and algorithmic complexity to assess potential copyright infringement. Degli Esposti, Lagioia, & Sartor's (2020) proposed approach involves analysing the frequency of n-grams (sequences of characters) in a text to define a similarity distance and / or quantification of 'character level similarity' (CLS: comparing the sequence of characters in an AI-generated work to those in the training corpus, aiming to ensure that the AI is generating new content that reflects the author's style while minimizing self-plagiarism) to measure the originality of AI-generated texts, aiming to provide quantitative support for copyright laws in the age of AI. Interestingly, Degli Esposti, Lagioia, & Sartor's (2020) suggest that even the concept of "style" though ambiguous may be approached mathematically.

Epstein et al. (2023b) observe that generative AI, with its ability to produce art, music, and literature, will radically transform the creative processes of artists. Skrodzki (2019) cites the keynote address of artist Mattis Kuhn in the German Conference on Artificial Intelligence proposing a taxonomy for understanding AI projects in the arts, including categories such as AI as creator, tool, subject, and medium showcasing Anna Ridler's artwork 'Myriad (Tulips)', where Ridler trained a machine learning algorithm on a dataset of 10,000 tulip images, underscoring the role of human labour in data labelling for AI art generation. Amato, et al. (2019) highlight the rise of 'pastiche' art, where AI technologies, like those in Elgammal, Liu, Elhoseiny, Mazzone's (2017) work, learn from existing artworks to create new pieces that imitate specific artistic styles.

Notaro (2020) notes that algorithmically derived art is not a new genre as Roman Verostko and the Algorists were an early 1960s group of visual artists that designed algorithms to generate art and considers the availability of Generative Adversarial Networks (GANs), along with an explosion in publicly available data to have democratised AI-generated creative practices.

Epstein, Levine, Rand, & Rahwan, (2020) observe that, despite a growing body of artistic and legal scholarship, public understanding of who or what an AI artist is, remains unclear. Hassine & Neeman (2019) critique projects such as the Dutch Next Rembrandt, the DEEPART project, and Obvious Art as examples of AI art that employ artistic plagiarism to produce art that is culturally biased. Epstein et al. (2020) highlight how public perception about AI anthropomorphism can be manipulated through the language used to talk about AI and report that already many AI-artists work in a bricolage style subject to the uncredited use of open-source software written by other artists.

Roose (2022) reports in the New York Times the curious case of Mr Jason M. Allen, an amateur artist who won the first prize at the Colorado State Fair competition by submitting an image generated by Midjourney. Such cases of course abound these days and don't necessarily make the news. However, the case of Allen raises concerns about plagiarism and how poetry and art plagiarism may be defined in the 21[st] century. Although as Roose reports artists defending Allen suggest that using AI to create art is analogous to using tools like Photoshop, since AI still requires human creativity in crafting prompts and guiding the process. Again, here I note the 'weak AI' narrative as a justification for using AI.

**5.5. Sustainability**

Finally, one needs to examine the environmental sustainability of AI art practices. Jääskeläinen, Pargman, & Holzapfel (2022) after noting the lack of comprehensive research quantifying the precise energy consumption and carbon footprint of AI art assert the need to study AI art's environmental footprint to understand its resource use and potential impact on sustainability and advocate for incorporating sustainability considerations into AI art research and artistic practices such as Life Cycle Assessment (LCA). Jääskeläinen, Pargman, & Holzapfel (2022) emphasise the need for artists to be aware of the environmental consequences of their AI-driven practices. Echoing the concerns of Devine (2019) who researched the political ecology of music and whose calculations demonstrate a lack of emissions reduction in the shift from physical materials to digital music streaming – often framed as 'dematerialisation' – even before factoring in the increasing use of AI technologies, which has the potential to further escalate the situation. Jääskeläinen, Pargman, & Holzapfel (2022) propose a framework for analysing AI art's environmental impact through two dimensions: materials (AI technologies: encompassing the hardware (GPUs, CPUs, servers) and software (online tools, installable software, programming libraries) used in AI art creation, the choice of which significantly influences energy consumption) and practices (artistic process, including phases like ideation, definition, actualisation, and display highlighting the significance of iterations in AI art, where repeated algorithmic runs can have a considerable energy impact despite minimal effort from the artist.). Complicating matters further is the need to define boundaries for peripheral activities included in the evaluation to assess AI art's full life cycle. Jääskeläinen, Pargman, & Holzapfel (2022) study shows that measuring the energy consumption of VQGAN+Clip, a popular AI art tool, on a laptop for a single inference iteration consumes 0.025-0.031 kWh over 3 hours, comparable to a small household appliance. Such energy consumption is further exacerbated by the variability in AI art's environmental impact, noting that some artists utilise massive training datasets, significantly increasing energy consumption. For instance, training the Bidirectional Encoder Representations from Transformers (BERT: a state-of-the-art natural language processing (NLP) model developed by Google) model can consume as much electricity as a trans-American flight.  Echoing such sustainability concerns Holzapfel, Jääskeläinen, & Kaila (2022:2) emphasizing the environmental impact of AI technologies particularly so energy consumption of training AI models on massive datasets reference the example of known AI-artist Refik Anadol who used *'45 terabytes of data – 587,763 image files, 1,880 video files, 1,483 metadata files, and 17,773 audio files equivalent of 40,000 hours of audio'* to train an AI model for his work raising significant concerns on the energy consumption and sustainability of such practices.

In the end, Jääskeläinen, Pargman, & Holzapfel (2022) propose exploring alternative aesthetics that value efficiency and resource consciousness drawing inspiration from artistic movements that thrive within limitations and are community-led, like the Nordic Demoscene, to foster environmentally conscious AI art practices.

Consequently, as Zylinska (2020) notes navigating the complexities of AI art requires an interdisciplinary dialogue encompassing the nature of creativity, the social, cultural, environmental and human impact of AI-generated content, and the ethical considerations surrounding AI development and use.  Historical and cultural context is crucial for understanding and evaluating the significance and impact of AI Art today especially since cultural differences influence attitudes toward AI-generated content (Cetinic & She, 2022). The undeniable reality is that already the rise of digital and AI-generated art in for example CryptoArt and NFTs but also in the adoption of collaborations between museums and cultural institutions with AI-artists (e.g. Mario Klingemann, Sougwen Chung, Refik Anadol) is transforming the art market and ownership dynamics as well as the environment and ethics. AI art forces us to reimagine the possibilities of creative expression in a technologically mediated world (Cetinic & She, 2022).

## 6. Human-AI Poetry Collaboration: Use Cases

David Young (2025), AI artist belonging to the AIArtists.org community, claims that the remarkable ability of AI technologies is their 'unique vision' compared to a normal human one and therefore AI should be understood as a non-human colleague who helps the artists to craft the works and assists in the material production of artworks. In this scenario (again focusing on the 'weak AI' narrative), the AI artist works collaboratively with an AI Engineer and various AI technologies to create the art.

Other artists such as Hannu Töyrylä express the same stance in viewing AI as a material and tool in artmaking (Slotte Dufva, 2023).  Töyrylä views AI as an augmentative tool, an active participant, enabling experimentation and exploration, allowing the artist to follow their path and direction using AI as an art-making material that actively participates in the art-making process.  Nevertheless, Töyrylä argues against AI as an autonomous artist since a level of experiential knowledge is necessary for sense-making (Slotte Dufva, 2023).

AI-generated poetry is distinct from poetry understood as an excess or economy of language, the work of a corporeal poet with emotions, experiences and passions (Strehovec, 2023). Poetry practice emphasises research, where the poet considers themselves in the role of researcher, producer and cognitive worker directing poetry-making as a sophisticated form of labour (Strehovec, 2023).

Understanding AI generated poetry is paramount, as a practice that will transform the way we engage with and create poetry (Oliveira, 2017). The use of AI in poetry leads to a reconsideration of the reader's experience, as the boundaries between human and AI-generated content become increasingly blurred (Lyu, Wang, Lin, & Wu, 2022).

Research on Human-AI co-creation reveals the importance of understanding how humans perceive and interact with AI during the creative process (Lyu, Wang, Lin, & Wu, 2022). While analysing the perceptual differences between Human-made and AI-generated art can shed light on the factors that influence our judgments of creativity and artistic value (Lyu, Wang, Lin, & Wu, 2022). Therefore, engaging with and understanding the cases of Human-AI collaboration is important especially, since as discussed in the previous sections public audiences are already unable to distinguish between Human-made and AI-generated poetry.

## 6.1. AI Systems for Poetry Generation / Collaboration

Humanity has always been fascinated with the potential for human augmentation and although most paradigms have failed to deliver on their promises occasionally with horrible practices and results (from the horrific practices of Hitler's Eugenics to the failure of Transhumanism to deliver on its promises with most start-ups embroiled in financial scandals). Similarly, AI technologies producing poetry are promoted as having the potential for AI to augment human creativity in poetry writing.

The simplest form of AI systems producing poetry encompasses experimental systems known as electronic assistants such as the Mamede, Trancoso, Araújo, & Viana (2004) system which focuses on poem classification and verse-ending word suggestions, using poetic concepts related to structure and metrics, along with grammatical categories and statistical language models, to predict suitable ending words. Or Xu, Miao, Chen, & Yang's (2020) RNN poetry generator working off Baidu user voice data and a relevant ancient Chinese poetry corpus to produce new poems; a system that can be used to train engineers' increasing comprehension of abstract theory and decreasing the fear of writing intelligent algorithms.

On the other hand, there are also fully functional and consumer ready systems such as the Verse-by-Verse system and of course LLMs like GPT-4, Claude and Gemini. Uthus, Voitovich, & Mical (2021) introduce the Verse-by-Verse system by Google that allegedly offers users the ability to compose a line of verse and then receive suggestions from a range of AI poets, each styled after a different established long-gone American poet. Uthus, Voitovich, & Mical (2021) suggest that such collaborative approaches aim to harness the strengths of both human and AI creativity, leading to new and unexpected poetic forms. The platform however remains to the day highly restrictive offering a single verse at a time working off user preferences on 3 established long-gone American poets.

As Gervás (2013) notes research on creativity from a computational perspective, what could otherwise be termed as 'computational creativity' recognises Boden's (1990) foundational work, which posits that artificial intelligence can elucidate creative thought through a conceptual space-search defined by constructive rules, while Sharples (1999) integrates this analysis with the understanding of writing as a problem-solving process that involves the writer as both a creative thinker and text designer. Interesting of course to note that such narratives reduce creativity to being or becoming a 'weak' faculty (see 'weak AI' in the previous sections) a tool that is, for a specific task.

As Manurung (2003) notes Gervás (2000) presents probably one of the first specialised systems for AI generated poetry the WASP system. WASP is a system trained in Spanish poetry constrained by poetic form (Romances, Cuartetos, Tercetos Encadenados) and allows users to iterate through drafts of poems, using reviser and judge modules to refine the form and content of the poem until it reaches a desired state. This iterative process, which mimics the way human poets often work, demonstrates how potentially humans may collaborate with AI. Manurung (2003) presents another specialised system by basing the system's development on evolutionary algorithms. The MCGONAGALL system which is partially successful in satisfying the conditions of grammaticality, meaningfulness, and poeticness demonstrates research that explores the creative possibilities of language, offering novel approaches to reimagining traditional poetic forms. However, as Manurung (2003) himself acknowledges, the limitations of such systems, particularly in accounting for discourse and complex semantic relationships, highlight the ongoing challenges in developing AI technologies that can fully grasp the nuances of human language.

An interesting example of work that explores the boundaries of Human-AI collaboration and in this case collaboration between AI, Robots and the public is the interactive art installation 'Dream Painter' whose purpose lies in representing in pictorial format via live painting the Collective Dream Space of the attending volunteering public. Thus, the installation allows the public to reflect on multiple interpretations of and engage with an embodied experience of their Collective Dream Space produced via oral dictation of their dreams (Guljajeva & Canet Sola, 2022).

Yang, Wang, & Wang (2024) suggest that the integration of AI into literary analysis may have the potential to change how we understand and interpret texts. Yang, Wang, & Wang's (2024) work on literary criticism suggest that traditional machine learning algorithms, like Support Vector Machines (SVM), can be used to classify literary works based on style and authorship, opening new avenues for research and pedagogy.

Weintraub and Correa-Díaz's (2023) extensive tome discusses Digital Poetics as a use case for AI in (digital) poetry in Latin America and showcases different ways artists are engaging with technology to create new forms of poetic expression. Weintraub and Correa-Díaz (2023) suggest that the analysis of digital poetics necessitates an understanding of both literary and computational concepts, highlighting the purely interdisciplinary nature of this emerging field.

Finally, Oliveira, Mendes, Boavida, Nakamura, & Ackerman (2019) develop the Co-PoeTryMe system. The Co-PoeTryMe system is an interactive poetry generation system enabling users to influence the output of the AI system by setting certain criteria and initial keywords if they want and explore new poetic possibilities. In this manner the Co-PoeTryMe system enables interactive poetry generation, allowing users to create poems through a co-creative process.

## 6.2. AI in Education

Fathoni (2023) suggests that AI rapidly transforming various fields amongst which education offers students potentially innovative tools for creation, exploration, and learning.

Vartiainen, Tedre, & Jormanainen (2023) discuss the use case of generative AI in Finnish K-9 education, particularly in creating digital art, providing opportunities for students to engage with AI technologies in a creative and allegedly meaningful way. Like authors cited previously Vartiainen, Tedre, & Jormanainen (2023) acknowledge that although AI-assisted tools can provide personalized feedback, help students explore different styles and forms, and encourage experimentation (Yang, Wang, & Wang, 2024), the integration of AI into the educational landscape requires striking a balance between leveraging AI's capabilities and fostering human creativity and critical thinking skills and hence a careful consideration of its pedagogical potential and the ethical implications of using AI with young learners.

Jiao (2022) discusses the use of AI technology in conjunction with probabilistic latent semantic analysis (PLSA) to analyse the semantic artistic conception of ancient Chinese Poetry as part of a multimedia-assisted artistic conception creation system. Jiao (2022) uses this system integrated in a VR/AR teaching environment alleging that the system has a significant positive impact in meeting the teaching needs by creating virtual characters and allowing learners to engage in a virtual scene, enhancing the immersive learning experience, therefore enhancing student understanding of the artistic conception of Ancient Chinese Poetry, thereby promoting their comprehensive literacy and language expression abilities. However, the system's performance, including its impact on immersion and teaching effectiveness, is at best nebulous since Jiao (2022) does not explain the immersion statistics.

Bedi (2023) explores the use of AI comics in education, suggesting that AI can enhance artistic expression among students aiding in the creation of comics by generating images, helping students focus on storytelling, narrative construction, and other creative aspects of the medium. Bedi (2023) astutely notes that the development of AI-literacy, encompassing skills and competencies in utilizing AI, is crucial for preparing individuals to engage with AI technologies critically, effectively and ethically. Such an approach encompassing understanding the ethical implications, potential biases inherent in AI technologies, and developing the ability to communicate and collaborate with AI in various domains may be the only way to educate students in using AI critically and reflectively (Bedi, 2023).

Fathoni (2023) highlights the fear among academics that students will use AI tools as shortcuts to complete assignments, potentially leading to plagiarism, fake authorship, diminished critical thinking, reduced knowledge and eventually rather disturbing cases of academic misconduct. Fathoni (2023) argues that to address these concerns, educators need to emphasise the importance critical thinking and transparency in the use of AI tools in education; advocating for assignments that require analysis, evaluation, and the combination of multiple concepts to encourage original thought and reduce the likelihood of plagiarism while encouraging students to be transparent about their use of AI, document their process, and explain how and why leveraging AI tools is crucial promoting in this manner accountability and ethical engagement with AI technologies.

Finally, Lin, Zecevic, Bouneffouf, & Cecchi (2023) discuss the use case of AI in therapeutic settings, where tools like 'Therapy View' utilise temporal topic modelling and AI-generated art to visualise therapy sessions, potentially aiding therapists in understanding patient narratives and facilitating therapeutic progress.

## 7. A Marketing Theory Critique

The use of Poetry in the field of Marketing and especially through the lens of CCT is an established method. Early work of pioneers in the area involved publishing poems in Marketing Journals that exemplified inquiries into various phaenomena (see e.g Schouten, 1993). In later works the use of Poetry evolved from academic observations to using poetry as a form of expression to gather qualitative data from research participants. The role of poetry in CCT and the theorisation of engagement with AI technologies from the lens of consumption studies offers insights as discussed below.

### 7.1. The Role of Poetry: Understanding AI narratives

Sherry & Schouten (2002) champion poetry as a research medium that cuts through the "crisis of representation" in consumer research. A crisis apparent in both AI technology development as well as the pervasive AI narratives. With the consumer and end-user being left out of any consultation in the development and marketing of AI technologies, Sherry & Schouten (2002) contend that conventional prose is inadequate for conveying the ineffable layers of consumer experience and identity. Poetry then functions as a powerful counter-discourse within the marketplace (Sherry & Schouten, 2002). The question of who consumes whom and who represents whom in current AI technologies remains looming. AI companies routinely employ marketing narratives to demystify complex algorithms — portraying AI products as possessing a "chain of thought" or exhibiting human-like creativity — such figurative marketing devices risk reducing nuanced human experiences to mere marketing clichés.

Tonner (2019) underscores that poetic language provides an 'emic' vocabulary which articulates intimate, often concealed consumption practices. This assertion contrasts starkly with AI marketers utilisation of sanitised, hyperbolic narratives. For instance, while most previous research emphasises that the potential of AI to generate art is both ethically challenging and aesthetically innovative, none of the papers reviewed sufficiently critique how figurative narratives — those same devices that might render AI "intelligent" in the public eye (strong AI narratives) — serve as a veneer over the technical limitations and reductive simplifications inherent in AI product marketing (weak AI narratives).

Following through Canniford's (2012) introduction of the 'Poetic Witness' offers methodological procedures—namely poetic transcription and poetic translation—to capture the multilayered, embodied realities of individual life worlds. Canniford (2012) demonstrates that a reflexive, poetically mediated research process yields insights into consumer experiences that remain inaccessible to traditional analytic methods. Hence, in contrast to previous research poetic methods can serve as a critical tool in deconstructing the market's fetishisation of AI technology. Poetic witness, therefore, becomes a means not only of representing but also of contesting the figurative narratives of perfection and fluid creativity that AI marketers promulgate.

Discussing figurative marketing narratives through the use of metaphor and simile as in the case of anthropomorphising AI technologies Hirschman's (2007) anthropological exploration of metaphor in the marketplace provides a valuable counterpoint, arguing that metaphoric imagery, far from being an innocuous stylistic choice, constructs deep symbolic meanings that both guide and restrict consumer behaviour. Hirschman's (2007, 2002a, 2002b) extensive work on metaphors and marketing narratives shows for example that in hair care products, metaphors cast these products as living entities requiring nourishment, thereby evoking cultural images of nature and vitality. Carefully crafted AI narratives like Anthropomorphising when co-opted by AI companies, transform abstract and black-box technical processes into relatable, almost organic phenomena. Yet this transformation, while effective in eliciting consumer trust, glosses over the dissonance between algorithmic 'reasoning' and genuine human creativity.

Brown and Wijland (2018) further dismantle the assumption that all figurative devices operate as minor embellishments to literal communication, arguing that simile and metonymy possess intrinsic analytical power, capable of revealing the underlying cognitive structures through which companies guide consumers to make sense of market offerings. Hence, when AI companies claim AI systems possess a 'chain of thought' or use language suggestive of human emotion, AI companies are, in effect, appropriating these rich figurative traditions to construct a narrative of sophistication and intimacy. This narrative, however, is inherently reductive—compressing complex opaque technological processes into neat, marketable soundbites that mask the algorithmic rigidity beneath.

The deployment of metaphoric language by AI companies, which often positions AI systems – products as having innate, human-like creative capacities, is a double-edged sword. On one hand, the metaphor simplifies consumer understanding of complex systems; on the other, risks engendering unrealistic expectations and obfuscating the material and computational limitations of AI technologies. As Nagl-Docekal and Zacharasiewicz (2022) note engaging with the pervasive public AI narratives and discussions one needs to fully interrogate how these metaphorical marketing narratives serve commercial ends and exhibit clear profit motives. Invoking metaphors of thought, creativity, and even emotion, AI companies are crafting a marketing narrative that is at odds with the necessary ambivalent, critically engaged perspectives of the AI market.

A further discussion on metaphors and corporate narratives is out of the scope of this review but I note here that in marketing theory metaphor analysis usually through a semiotic lens has a long and comprehensive tradition.

**7.2.    Poetic Agency and the Creator**

Wijland (2011) conceptualises poetic agency as the capacity of poetic language to reconfigure the meanings of consumer experiences.  In extension a 'good' or 'high-quality' poem may be one that asides grammaticality and poeticness has the ability to reconfigure the meaning of the audience's experience. Wijland (2011) illustrates that poetic agency does more than simply articulate experience — poetry actively disrupts and reimagines established narratives.  Poetic agency then is a form of resistance against the overly optimistic, deterministic and commodified narratives that characterise AI system – product branding. This resistance is particularly salient when one considers that AI companies, in their quest to humanise their technologies, employ metaphors that mirror poetic constructs, but without the critical depth that genuine poetic inquiry entails. One needs ask: Can AI write a poem that subverts the system's narrative and reconfigures audience experiences?

The notion of poetic agency brings forward the role, shape and form of the creator / poet.  Huber, Hua, & Deasy (2023) introduce the notion of 'creativity scars' – a concept capturing the psychological harms that may permanently impair an individual's creative potential. Such 'scars' may stem from early negative experiences (e.g. a childhood of harsh criticism) or professional contexts (e.g. lack of supportive mentorship) and leave enduring wounds that inhibit creative expression (Huber, Hua, & Deasy, 2023) 'Creativity scars' challenge the often-romanticised view of creativity as an unbounded, ever-renewable resource. For humans creative endeavour is fraught with personal risk often marred by personal sacrifice and emotional distress.

Echoing these sentiments Yachina & Fahrutdinova (2015) suggest that creativity is not only a process of artistic output but also one of personal formation.  Yachina & Fahrutdinova (2015) dissect the developmental trajectories of creative individuals underscoring that genuine creative expression involves overcoming deep-seated challenges. The creative individual is formed through a long, arduous process of personal and educational development. Hence creativity is a deeply embodied process — a process that generative AI technologies cannot replicate.

AI models may mimic creative outputs but do so without engaging with the profound experiential aspects that define human creativity. This divergence raises important questions about the authenticity and sustainability of AI-generated poetry when the algorithmic processes fail to account for the lived, often scarred, realities of the human creative journey. The scars borne by human creators, as painful as they may be, contribute to an artistic authenticity that is absent from algorithmically produced poems. Then if the emotional and experiential journey of human creators is essential to the aesthetic and cultural significance of poetry, AI system capacity to generate 'poetry' without these lived experiences results in outputs that may be technically proficient yet ultimately hollow. Even if the audience cannot distinguish between AI-generated poems and Human-made poems.

## 7.3. Consumer – AI Relationships: Authenticity, Disenchantment, and the Illusion of Choice.

### 7.3.1. Authenticity

The review delineates the transformative potential of AI in reshaping poetic production and artistic practice. However, a critique of Consumer – AI relationships in the marketplace, demonstrates an ambivalent picture — one in which the enchanting rhetoric employed by AI marketers masks profound consumer tensions.

Jago (2019) demonstrates that consumer perceptions of authenticity are intricately linked to the moral dimensions of the producer. Jago's (2019) results reveal that consumers attribute a lower degree of moral authenticity to algorithmic outputs compared to human work (see also section 4.2 for Algorithmic Aversion). In four between subjects experiments Jago (2019) shows that respondents rate algorithmic work as less ethically genuine — even when the technical accuracy (or type authenticity) is comparable to that of human endeavours.

Carroll & O'Connor (2019: 95) embracing Jago's (2019) argument note that *'the framing, or story matters a lot about how individuals interpret the meaning of authenticity'* and further report that if the stories of AI technology development and AI technology assumptions are explicitly, transparently and openly shared with public, then some AI systems would seem more authentic than others. Interestingly *'even anthropomorphizing actual hurricanes seems to affect responses to them, such as when individuals tend to prepare less for, and suffer more from, hurricanes with female compared with male names'* (Carroll & O'Connor (2019: 96). The dangers then of feminised anthropomorphising of AI systems (see section 5.3) are multi-fold.

In a market where AI products are increasingly marketed as embodying human-like creativity and ethical sensibilities, the reality suggests a disjunction between narrative and substance. This divergence is critical revealing that the enchanting metaphors deployed by AI companies are nothing more than a marketing artifice, one that ultimately fails to capture the genuine, multifaceted experience of consumer identity and agency suggesting that the very processes underpinning AI output engender scepticism among consumers. While AI companies celebrate the aesthetic and reflexive potentials of AI systems the marketplace's uncritical adoption of AI technologies may, in practice, erode consumer trust and diminish the perceived value of creative outputs.

### 7.3.2. Disenchantment and Enchantment

On the value of AI outputs for consumers and their engagement with AI technologies Fox's (2016) exploration of domesticating AI shows that the relationship between Consumers and AI is one of prosumption in an attempt to expand human self-expression, suggesting that technology is domesticated through a process of cultural negotiation rather than imposed as an external force. However, this simultaneous consumption and production not only of AI outputs but also AI narratives is often overlooked and downplayed in the pervasive narratives. AI companies have yet to recognise the contribution of users and creators as producers in the development and training of AI systems. Fox's (2016) analysis reveals that the enchantment derived from AI technologies — the promise to liberate self-expression and overcome production limitations — is ambivalent. The same AI technologies that enable unprecedented creativity simultaneously render consumers vulnerable to disillusionment, a precarious promise. The enchantment is contingent on cultural domestication; once the initial wonder fades, consumers may find themselves facing the cold logic of automated systems.

Belk, Weijo & Kozinets (2020) echo this disenchantment and develop the Disenchanted – Enchantment Model (DEM) articulating how consumer desire is initially captured through the seductive narratives of technology adoption, only to later give way to disenchantment as the limitations of these technologies become apparent. Whereas the reviewed articles contemplate AI technology's capacity to disrupt and reconfigure artistic production the very enchantment that propels technology adoption is inherently transient. The allure of a human-like creativity, emotion or logic may ultimately lead to consumer scepticism, as the novelty wears off and the disjunction between technological promise and lived experience becomes stark. The DEM framework serves as a potential to counter the overly optimistic narratives in AI marketing; recognising that the cultural capital of AI technologies is subject to cycles of fervour and disillusionment.

### 7.3.3. The Illusion of Choice

In a parallel path with disenchantment and potentially a factor of disenchantment is the existence of meaningful choices in the market, something that is currently problematic given the sterile nature of most AI systems. Anker (2023) interrogates the notion of choice within the burgeoning proliferation of generative AI models. In a market inundated with alternatives, consumers are often compelled to make choices that, upon closer scrutiny, lack substantive meaning. Anker (2023) argues that such choices are "existential" in nature only insofar as the choices provide a fleeting sense of agency; however, this impression is an illusion, as the overwhelming ubiquity and homogeneity of AI options trivialises the decision-making process. This insight critically challenges the assumption that increased availability equates to enhanced consumer autonomy. The commodification of choice both in terms of technologies as well as outputs of said technologies may ultimately reduce the consumer experience to a series of meaningless selections, thereby undermining the possibility of genuine self-expression and existential fulfilment.

While AI technologies can enhance convenience, personalization, performance and efficiency, AI deployment also engenders tensions related to privacy, control and alienation contributing to engagements that contrary to pervasive narratives do not contain only benefits but also inherent costs (Puntoni, Walker Reczek, Giesler, & Botti, 2020). Consumer experiences are not monolithic; rather, they encompass a spectrum of emotional, cognitive and social dimensions contrasting with the predominantly celebratory tone of pervasive narratives which tend to focus on the potentialities of AI technologies in expanding creative horizons. Puntoni et al. (2020) remind us that the same technologies that facilitate prosumption and self-expression can also contribute to a sense of loss — loss of control over personal data, loss of authentic human interaction, and ultimately, a diminished sense of agency.

In the Consumer – AI relationship technologies mediate consumption in ways that are inherently opaque and recursive, blurring the lines between consumer empowerment and corporate control (Airoldi and Rokka, 2022). Algorithmic agencing is not innocuous. While AI-generated poetry offers novel forms of expression, AI-generated Poetry also participates in the commodification and marketisation of creativity. Following through Wei and Geiger (2024) focusing on the agentic role of algorithms in shaping market dynamics contend that algorithms, far from being neutral tools, actively construct consumer experiences and influence market outcomes — an assertion that complicates the optimistic view of AI's creative capabilities.

Synthesising these perspectives, becomes clear that the current Consumer – AI relationships are paradoxical. On one side, AI technologies are celebrated for their ability to expand the horizons of creative self-expression and offer novel, personalised experiences; on the other, the very processes that enable these outcomes simultaneously impose a regime of control, surveillance and superficial choice that risks disenchanting and alienating consumers.

### 7.4. Poetry and Art as Consumer Goods

Drummond (2006) traces the evolution of Caravaggio's oeuvre from the historical origins in a sacred, patronage-driven system to the painter's contemporary status as a mass-consumed commodity, presenting a compelling historical case: how a once-revolutionary artistic vision is recontextualised and diluted as the vision and the works become a commodity for mass consumption. Drummond (2006) proposes a five-phase model — Creation, Quotation, Interpretation, Recontextualisation, and

Consumption — and demonstrates how the intrinsic value and aura of art are progressively diluted as art moves through various stages of commodification. Drummond (2006) illustrates that the value of art is not merely that of aesthetic appeal but also that of the ability to convey rich, culturally embedded narratives that resound on a deeply personal level.

When the unique voice of the creator is replaced by AI's algorithmically mediated output, what is lost is the 'soul' of the art — a nuance that consumers often subconsciously value. While AI technologies may offer a seductive shortcuts to creative production, the same AI technologies simultaneously risk erasing the individual voices of creators — transforming art into a mere marketised commodity. This transformation not only diminishes the artistic quality of the work but also contributes to a broader trend of consumer disillusionment, where the promise of personalised, meaningful experiences is undermined by the homogenising logic of mass production.

Combined with the abundance of choices in the AI technology landscape where every AI system generates via a similar algorithmic process poetic or artistic outputs, consumer choice is reduced to a superficial ritual, devoid of substantive differentiation or authentic creative input (Anker, 2023). This phenomenon is analogous to the emergence of "Hotel Art" – art commissioned for commercial environments that, while functional for branding purposes, often lacks the vibrancy and originality that distinguishes truly creative works. The implication for consumers and poets/artists is stark: while generative AI technologies may offer cost efficiencies and scalability, the overreliance on AI technologies risks homogenising creative expressions and eroding the distinctive cultural capital that poetry and art provide.

Venkatesh & Meamber (2006) reminds us that art, like any other commodity, is subject to the forces of market dynamics. Yet, unlike other products, art is also imbued with symbolic and affective meanings that transcend utilitarian functions. Hence, when poets and artists leverage generative AI technologies to produce poetry and art in scale the risk is that the output will be stripped of its deeper cultural and emotional resonance. Such outputs may not only fail to capture the audience's imagination but also contribute to a broader trend of devaluing Human creative work.

### 7.5. Criteria for Evaluating Poetry

Expanding the quantitative and algorithmic criteria for evaluating poetry this section focuses on discussing qualitative criteria that may serve as criteria for poetry and potentially AI output evaluation. Beyond grammaticality, meaningfulness, poeticness aesthetics, surprise and style that may potentially be solved and calculated algorithmically AI-generated poems should also be evaluated on artistic concentration, embodied experience, discovery, conditionality and narrative truth, and transformative potential (Faulkner, 2007). While AI systems should be evaluated on flexibility and mutability. 'Artistic Concentration' evaluates the degree to which a poem consistently focuses on a central theme or argument without veering into superfluous language, focusing on the craft of poetry, including attention to detail such as titles, lines, punctuation, sound, rhyme, figurative language, and word choice, a penetrating, unified, and focused approach to language that is also permeable and open (Faulkner, 2007). 'Embodied Experience' evaluates the degree to which a poem evokes sensory and emotional responses that allow the reader to viscerally connect with the poem's insights, creating an experience for the reader through vivid imagery and emotional engagement, allowing them to experience emotions and feelings in situ (Faulkner, 2007). 'Discovery' evaluates the degree to which a poem challenges conventional interpretations and offers unexpected insights that prompt a re-evaluation of established paradigms, teaching us to see something familiar in new or surprising ways, offering insights into ourselves and the human condition. 'Conditionality and Narrative Truth' evaluate the degree to which a poem must faithfully convey the core messages of the poem's context in a manner that is both transparent and evocative, acknowledging the partiality of stories and the conditional nature of point of view (Faulkner, 2007). Finally, 'Transformative Potential' evaluates the degree to which a poem not only articulates personal opinions but also inspires the reader to reconsider their understanding of the poetic context providing new insights, perspectives, or advocating human change (Faulkner, 2007).

At the same time AI systems should be flexible and mutable to be able to adapt to different communities and situations, acknowledging that cultural conceptions of 'good' poetry evolve over time

(Faulkner, 2007). The quality of poetic representation should be judged not solely on poetry's aesthetic form but also on its capacity to convey nuanced, multi-layered meanings (Faulkner, 2005).

By establishing rigorous criteria for poetic quality, one can better discern between authentic artistic expression and commodified outputs produced to satisfy market demands (Faulkner, 2016). Such a critical stance is essential to resist the erosion of creative value and the imperative to generate poetry that prioritises efficiency over originality.

Finally, Lafrenière and Cox (2012) propose a meta-framework that underscores the necessity of integrating not only normative but also substantive, and performative dimensions into evaluative practices. Lafrenière and Cox (2012) contend that arts and poetry should be appraised not merely on aesthetic appeal but on the work's capacity to disseminate knowledge effectively, forcing us to interrogate whether the poetic outputs of AI systems genuinely contribute to the intellectual and poetic discourse or merely serve as decorative adjuncts to market-driven narratives.

## 8. Conclusions, Limitations and Future Research Directions

In the end the question of whether AI technologies are anything more than 'stochastic parrots' remains unanswered. Middleton (2024) discussing the 'stochastic parrot' critique of AI language models compares the way LLMs and AI technologies work with Erasure Poetry. Working off Srikanth Reddy's Voyager Middleton (2024) argues that much like AI technologies Erasure Poetry and more generally language and literary work rely on repetition in many forms: e.g. genre, intertextuality, allusion, and quotation. And much like AI technologies Erasure poetry works by slicing and splicing words together. Hence, while AI technologies use algorithms to select and combine text, erasure poets use their own judgment and creativity but both AI and erasure poets may uncover latent or hidden meanings in a text an element Middleton (2024) terms as "paragrams" a hidden or occult configuration of letters or words awaiting recognition.

All in all, the future of AI in poetry and art by extension is marked by some promise and a more than healthy dose of uncertainty and challenges. While AI technologies may offer new tools and possibilities for creative expression, technologies and practices alike raise substantial ethical and philosophical questions that require more than careful consideration. The ability of AI technologies to process vast amounts of data and generate text with remarkable fluency may open-up new avenues for exploring the boundaries of poetic language and form. However, remains crucial to approach AI technologies in poetry with a critical and discerning lens, acknowledging technological and algorithmic limitations while exploring technology's potential to expand rather than contract our understanding of poetry and the human condition.

The philosophical implications of AI-generated poetry extend beyond grammaticality, meaningfulness, and poeticness and mere aesthetics, raising questions about the nature of creativity, intentionality in artistic expression, and the value placed on human authorship (Chen, 2023). As AI continues to advance and produce more sophisticated poetic outputs, AI challenges our understanding of what constitutes 'genuine' poetry and the extent to which algorithmic systems can capture and convey human emotions and experiences (Strehovec, 2023).

Some authors (e.g. Epstein et al., 2023; Oliveira et al., 2019) consider the future of AI and poetry as holding immense potential for innovation, experimentation, and cross-disciplinary collaboration. Weintraub & Correa-Díaz (2023) echoing this assertion see AI in poetry, as having the potential to cultivate a more inclusive, diverse, and dynamic literary landscape that celebrates the interplay between human imagination and technological advancement. Yet, most of the reviewed sources acknowledge that navigating this uncharted territory requires not only a spirit of curiosity, but also critical reflection, and a strong and principled commitment to ethical considerations (Chen, 2023).

Ultimately, the impact of AI on poetry will depend on how society chooses to develop and utilise this technology, ensuring that technology serves to enhance and enrich human creativity, rather than undermining or supplanting it, as is usually the case.

The creative interplay between human ingenuity and technological advancements in AI continues to shape the landscape of poetry, faster than most regulators and the public can follow and in classical human fashion artistic practice has already embraced working with AI faster than academics can critically evaluate and reflect on AI outcomes and creative processes. Such a pace should prompt ongoing discussions about the nature of poetry and by extension art, authorship, intellectual property rights, creativity and the role of the human imagination in the digital age.

## 9. An Auto-Ethnographic Account: The Cherry on Top!

Ready for the Cherry on top?

If you have reached the end of this literature review and did not use a platform like SciSpace (at typeset.io) to get a TL;DR (too long did not read summary), first of all THANK YOU, second you guessed right this literature review is a collaboration between the Human Author and AI. In fact, as a proper academic and insolent know-it-all, I did not use one but four different AI systems. I describe and discuss the experience in the following paragraphs. And be prepared for a remarkably bewildering auto-ethnographic account.

Before I begin my auto-ethnographic account I offer the relevant context. I am an Assistant Professor in Marketing and before the public release of ChatGPT my interest in AI was exhausted by my engagement with all things sci-fi. A few papers I had read here and there suggested that AI development despite its famous moments like DeepBlue's win over Garry Kasparov and AlphaGo's win over Lee Sedol, was a field of specialised computer science and specialist algorithms.

ChatGPT's public furore sparked in me the idea to work with current LLMs to test their abilities. And an interesting conflagration of factors led me to poetry. You see a few months before ChatGPT became public I had the chance to publish in Greek my protoleum poetry collection translated as 'Green' in English (A translation that ChatGPT failed to produce.) The collection despite being a protoleum work took me more than a decade to write, compile, edit, line edit, and eventually publish. The advent of a publicly accessible platform like ChatGPT signalled a turning point in the information century whereupon for the first time humanity gained access to a much more dangerous and pervasive toy than ever before. My mind immediately conjured (Greece is famous for its rather disproportionately large publication of poetry collections per year – estimates range from 500 to 2000) millions of poetry collections being published all at once. Poetry seems to be a pervasive mode of expression in social media platforms often to the detriment of its quality. So, ChatGPT seemed like a dangerously convenient way to produce even more conflated and curated identities. Therefore, I decided to investigate its abilities. In the past two (2) years and two (3) months I have experimented extensively with various AI systems.

I presented part of this work for the first time last year in the Consumer Culture Theory (CCT) conference in San Diego and surprisingly (for me) fellow academics were interested in learning more about it. So, I continued working. This review is the third (3[rd]) work on the same topic. My second work an extension of the work presented in the CCT conference is: "Who Said It Best Machine or Man? Contemplating On 'The Long Road'" and was submitted and will be hosted by the Digital Poetry Expo 2025, which is organised and curated by Versopolis, a European platform of international poetry festivals.

I cannot yet share the details of that work since it's currently live, but I invite you to visit the Digital Poetry Expo 2025 and take part in the project.

I describe below the process that led to the creation of this review, the AI systems I used and my experiences thereof. I also provide some details I can share from my work currently hosted by Versopolis.

### 9.1. On Poetry Generation

Despite the accolades by the developers of various Poetry generation systems my experience suggests that most of these systems are quite incompetent at producing Poetry that encompasses grammaticality, meaningfulness, poeticness, a sense of surprise as well as take advantage of cultural nuances and language polysemy. I have tried most systems referenced in this review to generate poetry.

Verse by Verse for instance which has irrational constraints for an AI system is more of a toy than anything else and possibly a research instrument for Google. Most platforms have a rather hazy relationship with data privacy and what happens to the data generated is at least unclear even if one is to take statements about deleting chats and outputs craftily written in privacy statements by lawyers at face value. Verse by Verse in all my effort consistently failed to produce a single meaningful output even after pretty much copying one of my poems whole as a starting point. None of the verses the system produced could even remotely be used.

I also used Co-PoeTryMe (Oliveira et al., 2019). Now this system is interesting. The system specialises in producing poems mostly in traditional forms (e.g. sonnet) from either a starting poem of your own or simply some relevant seed words. The system has a maximum size of fifteen verses by fifteen syllables per verse while also incorporating a surprise index (from 0 to 4). Interestingly the system does not create a title for any of its creations. The poems can be edited on the system or downloaded in notepad format (.txt). Although, I cannot reveal all the experimentation details with Co-PoeTryMe as aforementioned. Even with a surprise index of 0 the system tends to create rather full of surprise poems which however do not always achieve grammaticality and straightforward meaningfulness although in certain traditions of the early to mid-20$^{th}$ century would definitely be considered high in poeticness. However, issues of cultural nuances and language polysemy remain difficult if impossible to track.

I have also used Anthropic's Claude through a third party chatbot application called Dante which allows for (two) 2 important modifications. First, because the Dante platform is used mainly by corporations to create customer service chatbots there is a distinct emphasis on data privacy. Which means that neither any training texts nor any results are visible to the public or for that matter to the companies providing the relevant models. In a way this fact satisfied my delusions of privacy. I thought accessing the various AI models offered in Dante mainly Antropic's and OpenAI's models through the API meant I preserved some scrap of privacy. Second is that you can develop a model trained on pre-specified data. These data can take various forms and in my case were an as much as possible exhaustive list of published papers in academic journals discussing AI and Poetry (a big chunk of the papers I reference in this review). I then asked Dante to parse the data using Anthropic's Claude Sonnet 3.5 and produce a poem. Although the process necessitated multiple iterations to engineer a detailed prompt Dante using Anthropic's Claude Sonnet 3.5 was able to create poems that satisfied grammaticality and surface level poeticness but was not strong on surprise or language polysemy. Interestingly, Dante also has a creativity index from 0 to 1 and the answers between the two vary significantly even using the same prompt. Here the bot also produces a title for its poems.

Finally, I have also used ChatGPT (Plus subscription) extensively to both edit as well as generate poetry. Neither of these two processes is as easy as their creators want to advertise. Sure, ChatGPT is particularly good in crafting rather generic and lukewarm marketing text content but creating poetry that satisfies all the conditions as I set above takes extremely specified and lengthy prompts which can only be optimised after several iterations and very detailed and piecemeal instructions regarding the exact tasks the bot needs to perform. Otherwise, the results range from nonsensical to simply trivial. However, after you spent several days, you can create poems that are almost indistinguishable from Human-crafted poems. I am noting here that for producing the work I presented in the CCT conference I produced one-hundred and seventy-five (175) pages of chat logs spending a total of two-hundred and forty point seven (240.7) hours. However, ChatGPT even in the subscription versions still suffers from various technical issues. Currently, the way ChatGPT works is by maintaining a sort of episodic memory that lasts for as long as the chat with the bot is active and automatically disconnects the chat session after six (6) hrs of inactivity. Still to the day ChatGPT's system as a platform suffers from lags related to internet connectivity and issues with different browsers. Which means that to guarantee stability one needs to work in a very structured manner and continuously keep a log of their chats with the bot. Starting to work with ChatGPT for Poetry generation I very quickly discovered that long chat sessions led to an output degradation, especially long sessions over the 6hr limit which can be kept live as long as you keep feeding the bot input showed extreme degradation. At the same time unless one engineers a detailed prompt and not just asks the bot to produce poetry without very specific guidance the bot tends to produce generic poems otherwise known in colloquial terms as 'Vanilla poems'. In my work, presented in Versopolis I provide the exact details of my experimental procedure. Undoubtedly working with ChatGPT to produce text that satisfies the criteria I set above takes a lot of hours of prompt engineering and iterations.

**9.2.    On the Process of Compiling this Literature Review**

In order to compile and write the present review I used a combination of Google's Notebook LM, Anthropic's Claude OPUS 200K via the Dante platform, SciSpace and ChatGPT. I used the first two to provide me with the bulk of the text. I used Scispace additionally to get article summaries, ask specific questions on specific papers and paraphrase certain stubborn sentences from different articles. Finally, I used ChatGPT-4 to polish the article, find alternative words, try to locate suitable journal (a failed task) as well as assess whether ChatGPT could figure out whether I had used AI to compile the review. I also used ChatGPT-o3-mini-high to produce bulk text specifically for the Marketing Theory critique part of the article. I did this because neither Anthropic's Claude OPUS 200K nor Google's Notebook LM can use reasoning to develop a critique of a provided text based on other provided texts.

I am noting here that for the literature search I did not depend on any AI platform. That is, I manually searched through Google Scholar and traditional Academic Journal Databases to locate the articles of interest. For the marketing critique I also crafted special sets of prompts to guide ChatGPT-o3-mini-high on the expected critique I wanted the model to produce.

In order to compile and write the review I first focused on producing bulk text from Google's Notebook LM, Claude's Anthropic via the Dante platform; both of these systems allow for generation of text based on sources the user provides them with. Although Dante could process all articles at once, Notebook currently has a limit of fifty files per bot so in order to fit all the articles I created two bots: One with fifty articles and a second one with thirty-five articles as eighty-five was the number of articles I started with. I crafted and tried to give both Dante and Notebook the same prompt and here is where the problems start. Notebook currently has an extremely limited number of words one can enter in the prompt line. To be exact that number is three-hundred thirty-three words. Dante on the contrary can handle approximately up to eight hundred words as a prompt command. While Dante regardless of the background model can handle a psychological profile and detailed guidance on the exact specifics on how to write, Notebook can't. Trying to input a psychological profile of the persona I wanted the bot to adopt Notebook got stuck and replied 'The system cannot produce a response right now. Try again later.' and it wouldn't be the first time. Dante also has the ability to accept instructions step by step and then with a similar episodic memory like ChatGPT (although the Dante Platform automatically saves all chats and there is no temporary chat function) produces the final response to the acquired task. Notebook cannot the model gets stuck in some kind of infinite loop and keeps repeating 'I am still working on the feedback you gave me. I will print the updated response when I am ready'. I never managed to make Notebook produce the response with the step-by-step instructions I gave the bot. Claude Opus 200K although better in the aspects I just mentioned was no better in stickiness. I managed to get Claude Opus 200K stuck several times and I had to refresh the bot and re-write or change the prompt. However, despite all their problems both bots were able to eventually produce continuous, for the most part cohesive text of at least 1000 words. In fact, Notebook impressed me by being able to produce continuous and relatively cohesive text at the maximum size of 2000 words – including in-text citations. However, Notebook only did that once and I failed to re-produce the result no matter how many prompts and times I tried. Both bots were better than ChatGPT which can get stuck for a variety of reasons including momentary internet connection glitches. However, under the ChatGPT-o3-mini-high model I was able to produce lengthy coherent text which did actually perform as I instructed the bot.

Interestingly, the quality of the produced text after significant prompt engineering could reach what I would evaluate as an MSc level dissertation student or a first (1$^{st}$) year PhD student on the lower end of the scale. I was neither ecstatic nor happy with the results, but it was a text I could work on.

Current LLMs are like stubborn students who do not want to learn how to properly reference, stay within the word limits, and use generic sentences and passive voice. Not sure if I should be impressed, terrified or just saddened. My response was to laugh out loud, scaring my parents who still cannot fathom my job.

Generic sentences with absolutely no information were probably the foremost irritation of all. For example:

> 'AI's foray into poetry has sparked debate about its potential to understand and appreciate this art form. While AI can perform certain tasks traditionally associated with literary analysis, such as identifying themes, styles, and genre conventions through pattern and word recognition ('A. I. Richards Can Artificial Intelligence Appreciate Poetry.pdf', 1) (Notebook could not produce the correct citation and according to my instructions provided me with the filename of the article) *some argue that AI's ability to appreciate poetry is limited by its reliance on databases and algorithms.'*.

The number 1 in the previous example is not the page from the article Notebook based its sentence on but a weird and impractical citation system Google invented for reasons beyond my perceptual ability. The numbers are live hyperlinks that usually – not always – point to the text the bot has allegedly paraphrased. In most cases failing to understand why the bot had produced the sentence I had to open the article in my PDF reader and manually track the ideas in the article. Depending on the case I could do that faster if SciSpace could provide me with a relevant summary or answer my question or slower manually simply reading the article. When I asked Notebook LM to re-print the text using in-text APA style citations instead of live hyperlinked numbers, the bot – much like some of my PhD students – properly ignored me and went on to produce the text without numbers and without any additional citations. Thankfully printing the filenames of the articles seemed a better alternative than anything else.

All LLMs were particularly awful at creating citations and references despite the fact that all files allowed copying; however, deciding to do so all three (3) LLMs (Dante, Notebook, SciSpace) produced full citations and references. Contrary to ChatGPT-4 that hallucinates references even when provided with the original files. SciSpace and Anthropic's Claude Opus 200K face simple problems in that in a number of cases the systems simply could not parse the provided information and provide me with the correct reference even though that was available. Notebook LM was trickier with ChatGPT-4 undertones, the times the system decided not to provide me with references the problems ranged from hilarious like quoting myself (because I gave the bot a prompt explaining how I would like articles to be referenced using my name as example) to intriguingly odd as mixing the paper author's name with the authors cited in the paper as in basically creating misattribution. Other times, the mistakes were simple as just pulling in random a name from the original's article reference list and using that as the name of the author and occasionally creating sentences that completely and utterly misrepresented the paper, and the work reported therein. However, and this is important contrary to ChatGPT-4 that has the tendency to simply invent paper details and citations – in a fill-in the gaps – sort of heuristic, Notebook LM seems to simply use internal paper references as originals despite the fact that these papers were not to be found in their database. Interestingly in contrast to ChatGPT-4 which hallucinates references even when provided with the original article .pdfs ChatGPT-o3-mini-high produced the correct references and citations.

Another issue which I discovered almost too late was that I made the mistake in all my prompts to ask Notebook and Dante to provide me with at least 1000-word literature reviews for publication in a high-quality academic conference. By the time I had pulled together the text and was well into writing the review I realised that the number of references in the article did not match at all the number of sources I had provided both systems. After trying repeatedly and failing to make the bots produce reviews that contained all the provided references I decided to only focus on Notebook since of all the AI systems was the fastest in producing responses and create additional bots. The initial review compilation only referenced 45 articles almost half of what I had provided. Therefore, I had to go back and manually separate the articles that were used and not used in this first iteration and then create 4 more bots at 9-11 references per bot. This decision seemed to do the trick and allowed Notebook to produce text that contained all the provided sources. When I asked SciSpace to create a high-level summary of the sources I had provided the bot arbitrarily took the first 5 articles in the knowledge base I created and gave a summary based on those. No idea why. And I failed to make the bot produce anything else.

Comparatively between Claude OPUS 200K, Notebook, SciSpace, and ChatGPT-o3-mini-high Notebook was the fastest, then came SciSpace which however has a terrible UI and uploading the sources takes ages, then ChatGPT-o3-mini-high and then OPUS 200K. SciSpace was the worst in providing actual text you can just copy-paste but was useful in summarising articles and finding pieces of information in the text. Although in more than a few cases the search function of my PDF reader was markedly faster in locating specific phrases in the articles.

To the day the system that still provides in 95% of cases the most accurate and stylistically correct references is Google Scholar. Although, even Google Scholar has sometimes problems with news articles and website references.

In order to compile and write this version of the review that you have read, contrary to anecdotal accounts I have read online that suggest the text produced by bots is so low in quality that is essentially unusable I was able to copy-paste rather large swathes of text. However, and this is a big one, linking the various arguments from the various papers together in a coherent narrative that makes sense and critique the various papers and arguments I had to engage deeply with the papers read them, use SciSpace to summarise them again and again, ask multiple questions on each paper, and delete a vast amount of repetition inherent in bot text, including significantly re-writing generic statements, to the point that final result has absolutely no relationship to the starting point whatsoever in any shape or form, categorically.

In the end, I cannot evaluate the 'true' quality or usefulness of this review. Only You, the interested reader can do that.

# References

Abramson, D. (2022). AI's Winograd Moment; or: How Should We Teach Machines Common Sense? Guidance from Cognitive Science. *Artificial Intelligence and Human Enhancement: Affirmative and Critical Approaches in the Humanities*, *21*, 127.

Airoldi, M., & Rokka, J. (2022). Algorithmic consumer culture. *Consumption Markets & Culture*, 25(5), 411–428. https://doi.org/10.1080/10253866.2022.2084726.

Alowedi, N. A., & Al-Ahdal, A. A. M. H. (2023). Artificial Intelligence based Arabic-to-English machine versus human translation of poetry: An analytical study of outcomes. *Journal of Namibian Studies: History Politics Culture*, *33*, 1523-1538.

Amato, G., Behrmann, M., Bimbot, F., Caramiaux, B., Falchi, F., Garcia, A., Geurts, J., Gibert, J., Gravier, G., Holken, H. and Koenitz, H., (2019). AI in the media and creative industries. *arXiv preprint arXiv:1905.04175*.

Anker, T. B. (2023). Meaningful choice: Existential consumer theory. *Marketing Theory*, 24(4), 591–609. https://doi.org/10.1177/14705931231207317.

Badura, M., Lampert, M., & Dreżewski, R. (2022). System supporting poetry generation using text generation and style transfer methods. *Procedia Computer Science*, *207*, 3310-3319.

Beals, K. (2018). " Do the New Poets Think? It's Possible": Computer Poetry and Cyborg Subjectivity. *Configurations*, *26*(2), 149-177.

Bedi, K. (2023, May). AI comics as art: scientific analysis of the multimedia content of AI comics in education. In *2023 46th MIPRO ICT and Electronics Convention (MIPRO)* (pp. 750-753). IEEE.

Belk, R., Weijo, H., & Kozinets, R. V. (2020). Enchantment and perpetual desire: Theorizing disenchanted enchantment and technology adoption. *Marketing Theory*, 21(1), 25–51. https://doi.org/10.1177/1470593120961461.

Brown, S., & Wijland, R. (2018). Figuratively speaking: of metaphor, simile and metonymy in marketing thought. *European Journal of Marketing*, *52*(1/2), 328-347.

Brown, T., Mann, B., Ryder, N., Subbiah, M., Kaplan, J.D., Dhariwal, P., Neelakantan, A., Shyam, P., Sastry, G., Askell, A. and Agarwal, S., 2020. Language models are few-shot learners. *Advances in neural information processing systems*, *33*, pp.1877-1901.

Canniford, R. (2012). Poetic witness: Marketplace research through poetic transcription and poetic translation. *Marketing Theory, 12*(4), 391–409. https://doi.org/10.1177/1470593112457740

Carroll, G. R., & O'Connor, K. (2019). Comment on "Algorithms and Authenticity" by Arthur S. Jago. *Academy of Management Discoveries*, *5*(1), 95-96.

Cave, N. (2023). Nick Cave - The Red Hand Files. *The Red Hand Files*. https://www.theredhandfiles.com/chatgpt-making-things-faster-and-easier/

Cetinic, E., & She, J. (2022). Understanding and creating art with AI: Review and outlook. *ACM Transactions on Multimedia Computing, Communications, and Applications (TOMM)*, *18*(2), 1-22.

Chen, M. (2023). Take a Chance on Me: Aleatory Poetry, Generative AI, and the External Demarcation Problem. *The Journal of Aesthetics and Art Criticism*, *81*(4), 508-524.

Chung, J. J. Y. (2022, October). Artistic user expressions in AI-powered creativity support tools. In *Adjunct Proceedings of the 35th Annual ACM Symposium on User Interface Software and Technology* (pp. 1-4).

Chused, R. H. (2023). Randomness, AI art, and copyright. *Cardozo Arts & Entertainment Law Journal*, *40*(3), 621–656.

Clements, W. (2016, July). Poetry beyond the Turing test. In Electronic Visualisation and the Arts. BCS Learning & Development. http://dx.doi.org/10.14236/ewic/EVA2016.42.

Colton S., Goodwin, J., & Veale, T. (2012). Full FACE poetry generation. *In Proceedings of 3rd International Conference on Computational Creativity, ICCC 2012*, pages 95–102, Dublin, Ireland.

Colton, S., & Wiggins, G., A. (2012). Computational creativity: The final frontier? *In Proceedings of 20th European Conference on Artificial Intelligence (ECAI 2012), volume 242 of Frontiers in Artificial Intelligence and Application*s, pages 21–26, Montpellier, France. IOS Press.

Colton, S., Charnley, J., & Pease, A. (2011). Computational creativity theory: The face and idea descriptive models. *In Proceedings of the 2nd International Conference on Computational Creativity,* page 90–95, M´exico City, M´exico, April.

Degli Esposti, M., Lagioia, F., & Sartor, G. (2020). The use of copyrighted works by AI systems: Art works in the data Mill. *European Journal of Risk Regulation*, *11*(1), 51-69.

Demmer, T. R., Kühnapfel, C., Fingerhut, J., & Pelowski, M. (2023). Does an emotional connection to art really require a human artist? Emotion and intentionality responses to AI-versus human-created art and impact on aesthetic experience. *Computers in Human Behavior*, *148*, 107875.

Deng, Z., Yang, H., & Wang, J. (2024). Can AI Write Classical Chinese Poetry like Humans? An Empirical Study Inspired by Turing Test. *arXiv preprint arXiv:2401.04952*.

Derrida, J., & Spivak, Gayatri, C., (translator). (1976). Of Grammatology. *The Johns Hopkins University Press*. ISBN 10: 0801818419 / ISBN 13: 9780801818417

Devine, K. (2019). *Decomposed: the political ecology of music*. Cambridge, Massachusetts: The MIT Press.

d'Inverno, M., & McCormack, J. (2015). Heroic versus collaborative AI for the arts.

Elam, M. (2023). Poetry Will Not Optimize; or, What Is Literature to AI?. *American literature*, *95*(2), 281-303.

Elgammal, A., Liu, B., Elhoseiny, M., Mazzone, M., (2017)., CAN: Creative Adversarial Networks Generating "Art" by Learning About Styles and Deviating from Style Norms. *arXiv preprint arXiv:1706.07068*.

Epstein, Z., Hertzmann, A., Herman, L., Mahari, R., Frank, M.R., Groh, M., Schroeder, H., Smith, A., Akten, M., Fjeld, J. and Farid, H., Farid, H., Leach N., Pentland A. S., & Russakovsky, O., (2023a). Art and the science of generative AI: A deeper dive *(arXiv: 2306.04141). arXiv* [online]

Epstein, Z., Hertzmann, A., Investigators of Human Creativity, Akten, M., Farid, H., Fjeld, J., Frank, M.R., Groh, M., Herman, L., Leach, N. and Mahari, R., (2023b). Art and the science of generative AI. *Science*, *380*(6650), pp.1110-1111.

Epstein, Z., Levine, S., Rand, D. G., & Rahwan, I. (2020). Who gets credit for AI-generated art?. *Iscience*, *23*(9).

Fathoni, A. F. C. A. (2023). Leveraging generative AI solutions in art and design education: Bridging sustainable creativity and fostering academic integrity for innovative society. In *E3S Web of Conferences* (Vol. 426, p. 01102). EDP Sciences.

Faulkner, S. L. (2005, May). How do you know a good poem? Poetic representation and the case for criteria. In *Symposium conducted at the first international conference on qualitative enquiry* (Vol. 29, pp. 95-120).

Faulkner, S. L. (2007). Concern with craft: Using Ars Poetica as criteria for reading research poetry. *Qualitative inquiry*, *13*(2), 218-234.

Faulkner, S. L. (2016). The art of criteria: Ars criteria as demonstration of vigor in poetic inquiry. *Qualitative Inquiry, 22*(8), 662–665. https://doi.org/10.1177/1077800416634739.

Flick, C., & Worrall, K. (2022). The ethics of creative AI. In *The Language of Creative AI: Practices, Aesthetics and Structures* (pp. 73-91). Cham: Springer International Publishing.

Fox, S. (2016). Domesticating artificial intelligence: Expanding human self-expression through applications of artificial intelligence in prosumption. *Journal of Consumer Culture*, 18(1), 169–183. https://doi.org/10.1177/1469540516659126.

Galanter, P. (2020). Towards ethical relationships with machines that make art. *Artnodes*, (26), 1-9.

Gatys, L. A., Ecker, A. S., & Bethge, M. (2016). Image style transfer using convolutional neural networks. In *Proceedings of the IEEE conference on computer vision and pattern recognition* (pp. 2414-2423).

Gervás, P. (2000). WASP: Evaluation of different strategies for the automatic generation of spanish verse. In Proceedings of the AISB'00 Symposium on Creative and Cultural Aspects and Applications of AI and Cognitive Science, Birmingham, UK. AISB.

Gervás, P. (2013). Computational modelling of poetry generation. In *Artificial Intelligence and Poetry Symposium, AISB Convention* (Vol. 2, p. 2).

Grba, D. (2022). Deep else: A critical framework for ai art. *Digital*, *2*(1), 1-32.

Guljajeva, V., & Canet Sola, M. (2022, October). Dream painter: an interactive art installation bridging audience interaction, robotics, and creative AI. In *Proceedings of the 30th ACM international conference on multimedia* (pp. 7235-7236).

Gunser, V. E., Gottschling, S., Brucker, B., Richter, S., Çakir, D., & Gerjets, P. (2022). The pure poet: How good is the subjective credibility and stylistic quality of literary short texts written with an artificial intelligence tool as compared to texts written by human authors?. In *Proceedings of the Annual Meeting of the Cognitive Science Society* (Vol. 44, No. 44).

Harrison, S. H., Hua, M., & Deasy, D. (2024), Scar Tissue That I Wish You Saw: Creativity Scars and Their Impact on Creative Work, *Academy of Management Discoveries*, (ja), https://doi.org/10.5465/amd.2023.0141

Hassine, T., & Neeman, Z. (2019). The zombification of Art History: How AI resurrects dead masters, and perpetuates historical biases. *Journal of Science and Technology of the Arts*, *11*(2), 28-35.

Hirschman, E. C. (2002a). Dogs as metaphors: Meaning transfer in a complex product set. *Semiotica*, *2002*(139), 125-159. https://doi.org/10.1515/semi.2002.017.

Hirschman, E. C. (2002b). Metaphors, archetypes, and the biological origins of semiotics. *Semiotica*, *2002*(142), 315-349. https://doi.org/10.1515/semi.2002.076

Hirschman, E. C. (2007). Metaphor in the marketplace. *Marketing Theory*, *7*(3), 227-248.

Hitsuwari, J., Ueda, Y., Yun, W., & Nomura, M. (2023). Does human–AI collaboration lead to more creative art? Aesthetic evaluation of human-made and AI-generated haiku poetry. *Computers in Human Behavior*, *139*, 107502.

Holzapfel, A., Jääskeläinen, P., & Kaila, A. K. (2022). Environmental and Social Sustainability of Creative-Ai. *arXiv preprint arXiv:2209.12879*.

Horton Jr, C. B., White, M. W., & Iyengar, S. S. (2023). Bias against AI art can enhance perceptions of human creativity. *Scientific reports*, *13*(1), 19001.

Hutson, J., & Schnellmann, A., (2023). The Poetry of Prompts: The Collaborative Role of Generative Artificial Intelligence in the Creation of Poetry and the Anxiety of Machine Influence. *Faculty Scholarship*. 462. https://digitalcommons.lindenwood.edu/faculty-research-papers/462.

Jääskeläinen, P., Pargman, D., & Holzapfel, A. (2022, June). On the environmental sustainability of Ai art (s). In *Eighth workshop on computing within limits* (pp. 1-9).

Jago, A. S. (2019). Algorithms and authenticity. *Academy of Management Discoveries*, 5(1), 38–56. https://doi.org/10.5465/amd.2017.0002.

Jiang, H. H., Brown, L., Cheng, J., Khan, M., Gupta, A., Workman, D., Hanna, A., Flowers, J., & Gebru, T. (2023, August). AI Art and its Impact on Artists. *In Proceedings of the 2023 AAAI/ACM Conference on AI, Ethics, and Society* (pp. 363-374).

Jiao, Y. (2022). Research on the Artistic Conception of Multimedia-Assisted Ancient Poetry Based on AI Technology. *Scientific Programming*, *2022*(1), 8538503.

Jordan, J. (2020). AI as a Tool in the Arts. *Arts Management & Technology Laboratory at Carnegie Mellon University*. [accessed 16/01/2025 at: https://amt-lab.org/blog/2020/1/ai-as-a-tool-in-the-arts].

Karaban, V., & Karaban, A. (2024, January). AI-translated poetry: Ivan Franko's poems in GPT-3.5-driven machine and human-produced translations. In *Forum for Linguistic Studies* (Vol. 6, No. 1).

Kirke, A., & Miranda, E. (2013, April). Emotional and multi-agent systems in computer-aided writing and poetry. In *Proceedings of the Artificial Intelligence and Poetry Symposium* (pp. 17-22).

Kirmani, A. R. (2022). Artificial intelligence-enabled science poetry. *ACS Energy Letters*, *8*(1), 574-576.

Knochel, A. D., & Sahara, O. (Eds.). (2022). *Global media arts education: Mapping global perspectives of media arts in education*. Springer Nature.

Kobierski, M. (2023). AI the Creator? Analysing Prose and Poetry Created by Artificial Intelligence. *Currents*, *9*.

Köbis, N., & Mossink, L. D. (2021). Artificial intelligence versus Maya Angelou: Experimental evidence that people cannot differentiate AI-generated from human-written poetry. *Computers in human behavior*, *114*, 106553.

Lafrenière, D., & Cox, S. M. (2012). 'If you can call it a poem': Toward a framework for the assessment of arts-based works. *Qualitative Research, 1–19*. https://doi.org/10.1177/1468794112446104.

Li, X., & Zhang, B. (2020, October). AI poem case analysis: Take ancient Chinese poems as an example. In *Proceedings of the 2020 Conference on Artificial Intelligence and Healthcare* (pp. 132-136).

Lin, B., Zecevic, S., Bouneffouf, D., & Cecchi, G. (2023). TherapyView: Visualizing therapy sessions with temporal topic modeling and AI-generated arts. *arXiv preprint arXiv:2302.10845*.

Linardaki, 2022). Poetry at the first steps of Artificial Intelligence. *Humanist Studies & the Digital Age*, *7*(1).

Lyu, Y., Wang, X., Lin, R., & Wu, J. (2022). Communication in human–AI co-creation: Perceptual analysis of paintings generated by text-to-image system. *Applied Sciences*, *12*(22), 11312.

M. Boden, Creative Mind: Myths and Mechanisms, Weidenfeld & Nicholson, London, 1990.

Mamede, N., Trancoso, I., Araújo, P., & Viana, C. (2004). An electronic assistant for poetry writing. In Advances in Artificial Intelligence–IBERAMIA 2004: 9th Ibero-American Conference on AI, Puebla, Mexico, November 22-26, 2004. Proceedings 9 (pp. 286-294). Springer Berlin Heidelberg.

Manovich, L. (2018). AI Aesthetics. [Accessed online 04/01/2025 at https://www.karlancer.com/api/file/1733554475-mIVP.pdf ]

Manurung, H. (2003). An evolutionary algorithm approach to poetry generation. PhD Thesis. Institute for Communicating and Collaborative Systems School of Informatics. University of Edinburgh.

Mateas, M. (2001). Expressive AI: A hybrid art and science practice. *Leonardo*, *34*(2), 147-153.

Middleton, P. (2024). Parrots and Paragrams: AI Language Models and Erasure Poetry. Modern Philology, 121(3), 352-374.

Mikalonytė, E. S., & Kneer, M. (2022). Can artificial intelligence make art?: Folk intuitions as to whether AI-driven robots can be viewed as artists and produce art. *ACM Transactions on Human-Robot Interaction (THRI)*, 11(4), 1-19.

Mike Sharples, How We Write: Writing As Creative Design, Routledge, June 1999.

Miller, A. I. (2019). *The artist in the machine: The world of AI-powered creativity*. Mit Press.

Misztal, J., & Indurkhya, B. (2014, June). Poetry generation system with an emotional personality. In *ICCC* (pp. 72-81).

Murad, H., & Rahman, R. (2023). AI Poet: A Deep Learning-based Approach to Generate Artificial Poetry in Bangla. In *Applied Intelligence for Industry 4.0* (pp. 188-197). Chapman and Hall/CRC.

Nagl-Docekal, H., & Zacharasiewicz, W. (Eds.). (2022). *Artificial Intelligence and Human Enhancement: Affirmative and Critical Approaches in the Humanities* (Vol. 21). Walter de Gruyter GmbH & Co KG.

Newton, A., & Dhole, K. (2023). Is AI art another industrial revolution in the making?. *arXiv preprint arXiv:2301.05133*.

Notaro, A. (2020). State-of-the-art: AI through the (artificial) artist's eye. *EVA London 2020: Electronic Visualisation and the Arts*, 322-328.

Oliveira, H. G. (2017, September). A survey on intelligent poetry generation: Languages, features, techniques, reutilisation and evaluation. In *Proceedings of the 10th international conference on natural language generation* (pp. 11-20). Santiago de Compostela, Spain: Association for Computational Linguistics.

Oliveira, H. G., Mendes, T., Boavida, A., Nakamura, A., & Ackerman, M. (2019). Co-PoeTryMe: interactive poetry generation. *Cognitive Systems Research*, *54*, 199-216.

Phelan, J. (2021). " AI Richards": Can Artificial Intelligence Appreciate Poetry?. *Philosophy and Literature*, *45*(1), 71-87.

Piskopani, A. M., Chamberlain, A., & Ten Holter, C. (2023, July). Responsible AI and the arts: The ethical and legal implications of AI in the arts and creative industries. In *Proceedings of the First International Symposium on Trustworthy Autonomous Systems* (pp. 1-5).

Pretsch, E. (2023). Artificial Intelligence and creativity in poetry: effect of AI-written poems on human emotions. *Journal of Creativity and Inspiration. 1(1). (*[accessed online 16/01/2025 at: https://www.poeticmind.co.uk/journal-creativity-and-inspiration/volume-1-issue-1/artificial-intelligence-and-creativity-in-poetry-effect-of-ai-written-poems-on-human-emotions/].

Puntoni, S., Walker Reczek, R., Giesler, M., & Botti, S. (2020). Consumers and artificial intelligence: An experiential perspective. *Journal of Marketing*, 85(1), 131–151. https://doi.org/10.1177/0022242920953847.

Revell, G. (2022, September). Madeleine: Poetry and art of an artificial intelligence. In *Arts* (Vol. 11, No. 5, p. 83). MDPI.

Roose, K. (2022). An AI-Generated Picture Won an Art Prize. Artists Aren't Happy. *New York Times*. [16/01/2025 accessible via a paywall at: https://www.nytimes.com/2022/09/02/technology/ai-artificial-intelligence-artists.html].

Saurini, E., (2023). Creativity in Art and Academia: Analyzing the Effects of AI Technology Through the Lens of ChatGPT. *Regis University Student Publications* (comprehensive collection). 1102. https://epublications.regis.edu/theses/1102

Schouten, J. (1993). Sixty-five type one convertible. *International Journal of Research in Marketing, 10*(3), 339.

Sherry, J. F. Jr., & Schouten, J. W. (2002). A role for poetry in consumer research. *Journal of Consumer Research, 29*(2), 215–227.

Skrodzki, M. (2019). AI and Arts–A Workshop to Unify Arts and Science. *w/k-Zwischen Wissenschaft & Kunst – Between Science & Art Journal*. https://doi.org/10.55597/e5342.

Slater, A. (2023). Post-Automation Poetics; or, How Cold-War Computers Discovered Poetry. *American Literature*, *95*(2), 205-227.

Slotte Dufva, T. (2023). Entanglements in AI Art. In A. D. Knochel, & O. Sahara (Eds.), Global Media Arts Education (pp. 181-196). (Palgrave Studies in Educational Futures). Palgrave Macmillan. https://doi.org/10.1007/978-3-031-05476-1_11

Spencer, S. L. (2023). What It Means to Have Meaning: AI's Poetic Appropriation of the Human Imagination. *The Faculty of the College of Arts and Sciences. Liberty University.*

Srinivasan, R., & Uchino, K. (2021, March). Biases in generative art: A causal look from the lens of art history. In *Proceedings of the 2021 ACM Conference on Fairness, Accountability, and Transparency (FAccT'21)* (pp. 41-51).

Strehovec, J. (2023). From Poetry As the Excess of Language to Poems Generated by Artificial Intelligence (Poet As Researcher in the World of New Cultural Paradigms). *Canadian Journal of Language and Literature Studies*, *3*(4), 98-111.

Tanasescu, C., Kesarwani, V., & Inkpen, D. (2018, May). Metaphor detection by deep learning and the place of poetic metaphor in digital humanities. In *The thirty-first international flairs conference*.

Tonner, A. (2019). Consumer culture poetry: Insightful data and methodological approaches. *Consumption Markets & Culture, 22*(3), 256–271. https://doi.org/10.1080/10253866.2018.1474110

Tromble, M. (2020). Ask not what AI can do for art... but what art can do for AI. *Artnodes*, (26), 1-9.

Uthus, D., Voitovich, M., & Mical, R. J. (2021). Augmenting poetry composition with verse by verse. *arXiv preprint arXiv:2103.17205*.

Uthus, D., Voitovich, M., Mical, R., & Kurzweil, R. (2019). First steps towards collaborative poetry generation. In *NeurIPS Workshop on Machine Learning for Creativity and Design* (Vol. 3).

Vartiainen, H., Tedre, M., & Jormanainen, I. (2023). Co-creating digital art with generative AI in K-9 education: Socio-material insights. *International Journal of education through art*, *19*(3), 405-423.

Venkatesh, A., & Meamber, L. (2006). Arts and aesthetics: Marketing and cultural production. *Marketing Theory, 6*(1), 11–39.

Vyas, B. (2024). Ethical Implications of Generative AI in Art and the Media. *International Journal for Multidisciplinary Research (IJFMR), E-ISSN*, 2582-2160.

Wei, R., & Geiger, S. (2024). Algorithmic agencing in platform markets. *Marketing Theory*, 1–19. https://doi.org/10.1177/14705931241275558.

Weintraub, S., & Correa-Díaz, L. (Eds.). (2023). *Latin American Digital Poetics*. Springer International Publishing AG.

West, R., & Burbano, A. (2020). AI, arts & design: Questioning learning machines. *Artnodes*, (26), 1-9.

Wijland, R. (2011). Anchors, mermaids, shower-curtain seaweeds and fish-shaped fish: The texture of poetic agency. *Marketing Theory, 11*(2), 127–141. https://doi.org/10.1177/1470593111403217

Wu, C. C., Song, R., Sakai, T., Cheng, W. F., Xie, X., & Lin, S. D. (2019, September). Evaluating image-inspired poetry generation. *In CCF international conference on natural language processing and Chinese computing* (pp. 539-551). Cham: Springer International Publishing.

Xu, X., Miao, J., Chen, Y., & Yang, S. (2020, November). Poetry Automatic Generation System Based On Natural Language Processing. *In The 8th International Symposium on Test Automation & Instrumentation (ISTAI 2020)* (Vol. 2020, pp. 90-93). IET.

Yachina, N., & Fahrutdinova, G. (2015). Formation of the creative person. *Procedia-Social and Behavioral Sciences*, *177*, 213-216.

Yang, L., Wang, G., & Wang, H. (2024). Reimagining Literary Analysis: Utilizing Artificial Intelligence to Classify Modernist French Poetry. *Information*, *15*(2), 70.

Young, D., (2025)., Young on using artificial intelligence., [accessed 16/01/2025 at: https://aiartists.org/david-young].

Zulić, H. (2019). How AI can change/improve/influence music composition, performance and education: three case studies. *INSAM Journal of Contemporary Music, Art and Technology*, (2), 100-114.

Zylinska, J. (2020). AI ART: Machine Visions and Warped Dreams. Open Humanities Press.